\documentclass[aps,prl,preprint,tightenlines,superscriptaddress,showpacs,byrevtex]{revtex4}

\usepackage{graphicx} 
\usepackage{dcolumn}  

\graphicspath{{ps}}

\newcommand{\BF}{\ensuremath{{\cal{B}}}}
\newcommand{\Am}{\ensuremath{{\cal{A}}}}
\newcommand{\UFS}{\ensuremath{\Upsilon(4S)}}
\newcommand{\bbbar}{\ensuremath{B\bar{B}}}
\newcommand{\qqbar}{\ensuremath{q\bar{q}}}

\newcommand{\de}{\ensuremath{\Delta E}}
\newcommand{\mb}{\ensuremath{M_{\rm bc}}}

\newcommand{\mom}{GeV/$c$}
\newcommand{\mass}{MeV/$c^2$}
\newcommand{\Mass}{GeV/$c^2$}
\newcommand{\Masssq}{(GeV/$c^2$)$^2$}

\newcommand{\bcppp}{\ensuremath{B^+\to \pi^+\pi^+\pi^-}}
\newcommand{\bckpp}{\ensuremath{B^+\to K^+\pi^+\pi^-}}
\newcommand{\bckkk}{\ensuremath{B^+\to K^+K^+K^-}}
\newcommand{\bnkpp}{\ensuremath{B^0\to K^0_S\pi^+\pi^-}}

\newcommand{\Kpp}{\ensuremath{K\pi\pi}}

\newcommand{\kpp}{\ensuremath{K^+\pi^+\pi^-}}

\newcommand{\kspp}{\ensuremath{K^0_S\pi^+\pi^-}}

\newcommand{\pipi}{\ensuremath{\pi^+\pi^-}}

\newcommand{\kcpi}{\ensuremath{K^+\pi^-}}
\newcommand{\kspi}{\ensuremath{K^0_S\pi^{\pm}}}

\newcommand{\ks}{\ensuremath{K^0_S}}

\newcommand{\chic}{\ensuremath{\chi_{c0}}}

\newcommand{\sfs}{\ensuremath{s_{12}}}
\newcommand{\sft}{\ensuremath{s_{13}}}
\newcommand{\sst}{\ensuremath{s_{23}}}

\newcommand{\lumi}{\ensuremath{357~{\rm fb}^{-1}}}

\def\nima#1#2#3{{Nucl.\ Instr.\ and Meth.} {\bf A#1}, #3 (#2)}

\def\plb#1#2#3{{ Phys.\ Lett.}  {\bf B#1}, #3 (#2)}

\def\prd#1#2#3{{ Phys.\ Rev.}   {\bf D#1}, #3 (#2)}

\begin{document}

\preprint{\vbox{ \hbox{ \vspace*{10mm}  }
                 \hbox{BELLE-CONF-0577}
}}

\vspace*{10mm}

\title{ \quad\\[0.5cm] 
Dalitz analysis of the three-body charmless decay $\bnkpp$}

\affiliation{Aomori University, Aomori}
\affiliation{Budker Institute of Nuclear Physics, Novosibirsk}
\affiliation{Chiba University, Chiba}
\affiliation{Chonnam National University, Kwangju}
\affiliation{University of Cincinnati, Cincinnati, Ohio 45221}
\affiliation{University of Frankfurt, Frankfurt}
\affiliation{Gyeongsang National University, Chinju}
\affiliation{University of Hawaii, Honolulu, Hawaii 96822}
\affiliation{High Energy Accelerator Research Organization (KEK), Tsukuba}
\affiliation{Hiroshima Institute of Technology, Hiroshima}
\affiliation{Institute of High Energy Physics, Chinese Academy of Sciences, Beijing}
\affiliation{Institute of High Energy Physics, Vienna}
\affiliation{Institute for Theoretical and Experimental Physics, Moscow}
\affiliation{J. Stefan Institute, Ljubljana}
\affiliation{Kanagawa University, Yokohama}
\affiliation{Korea University, Seoul}
\affiliation{Kyoto University, Kyoto}
\affiliation{Kyungpook National University, Taegu}
\affiliation{Swiss Federal Institute of Technology of Lausanne, EPFL, Lausanne}
\affiliation{University of Ljubljana, Ljubljana}
\affiliation{University of Maribor, Maribor}
\affiliation{University of Melbourne, Victoria}
\affiliation{Nagoya University, Nagoya}
\affiliation{Nara Women's University, Nara}
\affiliation{National Central University, Chung-li}
\affiliation{National Kaohsiung Normal University, Kaohsiung}
\affiliation{National United University, Miao Li}
\affiliation{Department of Physics, National Taiwan University, Taipei}
\affiliation{H. Niewodniczanski Institute of Nuclear Physics, Krakow}
\affiliation{Nippon Dental University, Niigata}
\affiliation{Niigata University, Niigata}
\affiliation{Nova Gorica Polytechnic, Nova Gorica}
\affiliation{Osaka City University, Osaka}
\affiliation{Osaka University, Osaka}
\affiliation{Panjab University, Chandigarh}
\affiliation{Peking University, Beijing}
\affiliation{Princeton University, Princeton, New Jersey 08544}
\affiliation{RIKEN BNL Research Center, Upton, New York 11973}
\affiliation{Saga University, Saga}
\affiliation{University of Science and Technology of China, Hefei}
\affiliation{Seoul National University, Seoul}
\affiliation{Shinshu University, Nagano}
\affiliation{Sungkyunkwan University, Suwon}
\affiliation{University of Sydney, Sydney NSW}
\affiliation{Tata Institute of Fundamental Research, Bombay}
\affiliation{Toho University, Funabashi}
\affiliation{Tohoku Gakuin University, Tagajo}
\affiliation{Tohoku University, Sendai}
\affiliation{Department of Physics, University of Tokyo, Tokyo}
\affiliation{Tokyo Institute of Technology, Tokyo}
\affiliation{Tokyo Metropolitan University, Tokyo}
\affiliation{Tokyo University of Agriculture and Technology, Tokyo}
\affiliation{Toyama National College of Maritime Technology, Toyama}
\affiliation{University of Tsukuba, Tsukuba}
\affiliation{Utkal University, Bhubaneswer}
\affiliation{Virginia Polytechnic Institute and State University, Blacksburg, Virginia 24061}
\affiliation{Yonsei University, Seoul}
  \author{K.~Abe}\affiliation{High Energy Accelerator Research Organization (KEK), Tsukuba} 
  \author{K.~Abe}\affiliation{Tohoku Gakuin University, Tagajo} 
  \author{I.~Adachi}\affiliation{High Energy Accelerator Research Organization (KEK), Tsukuba} 
  \author{H.~Aihara}\affiliation{Department of Physics, University of Tokyo, Tokyo} 
  \author{K.~Aoki}\affiliation{Nagoya University, Nagoya} 
  \author{K.~Arinstein}\affiliation{Budker Institute of Nuclear Physics, Novosibirsk} 
  \author{Y.~Asano}\affiliation{University of Tsukuba, Tsukuba} 
  \author{T.~Aso}\affiliation{Toyama National College of Maritime Technology, Toyama} 
  \author{V.~Aulchenko}\affiliation{Budker Institute of Nuclear Physics, Novosibirsk} 
  \author{T.~Aushev}\affiliation{Institute for Theoretical and Experimental Physics, Moscow} 
% ...
\collaboration{The Belle Collaboration}

\begin{abstract}
    We report preliminary results of a Dalitz plot analysis of three-body
charmless $\bnkpp$ decays. The analysis is performed with a data sample that
contains 386 million $B\bar{B}$ pairs collected near the $\Upsilon(4S)$
resonance with the Belle detector at the KEKB asymmetric energy $e^+ e^-$
collider. Measurements of branching fractions for quasi-two-body decays
$B^0\to\rho(770)^0K^0$, $B^0\to f_0(980)K^0$, $B^0\to K^*(892)^+\pi^-$,
$B^0\to K^*(1430)^+\pi^-$, and upper limits on several other quasi-two-body
decay modes are reported.
\end{abstract}


\maketitle

{\renewcommand{\thefootnote}{\fnsymbol{footnote}}}
\setcounter{footnote}{0}

\section{Introduction}

     First results of amplitude analyses of $B$ meson decays to a several
three-body charmless hadronic final states have been reported recently:
$\bckkk$~\cite{belle-khh3}, $\bckpp$~\cite{belle-khh3,belle-khh4,babar-ppp},
$B^0\to K^+\pi^-\pi^0$~\cite{babar-kpp} and $\bcppp$~\cite{babar-ppp}.
Branching fractions for a number of quasi-two-body decays have been measured
with some of them being observed for the first time. 

In this paper we present preliminary results of a Dalitz plot analysis of
neutral $B$ meson decay to the $\kspp$ three-body charmless final state. The
analysis is based on a $\lumi$
data sample containing 386 million $B\bar{B}$ pairs, collected  with the Belle
detector operating at the KEKB asymmetric-energy $e^+e^-$ collider~\cite{KEKB}
with a center-of-mass (c.m.) energy at the $\Upsilon(4S)$ resonance
(on-resonance data). The beam energies are 3.5 GeV for positrons and 8.0 GeV
for electrons. For the study of the $e^+e^-\to q\bar{q}$ continuum background,
we use data taken 60~MeV below the $\UFS$ resonance (off-resonance data).


\section{The Belle detector}

  The Belle detector~\cite{Belle} is a large-solid-angle magnetic spectrometer
based on a 1.5~T superconducting solenoid magnet. Charged particle tracking is
provided by a silicon vertex detector and a 50-layer central drift chamber
(CDC) that surround the interaction point. Two inner detector configurations
were used. A 2.0 cm beampipe and a 3-layer silicon vertex detector was used
for the first sample of 152 million $B\bar{B}$ pairs, while a 1.5 cm beampipe,
a \mbox{4-layer} silicon detector and a small-cell inner drift chamber were
used to record
the remaining 234 million $B\bar{B}$ pairs~\cite{Ushiroda}. The charged
particle acceptance covers laboratory polar angles between $\theta=17^{\circ}$
and $150^{\circ}$, corresponding to about 92\% of the total solid angle in
the c.m.\ frame. The momentum resolution is determined from cosmic rays and
$e^+ e^-\to\mu^+\mu^-$ events to be $\sigma_{p_t}/p_t=(0.30\oplus0.19p_t)\%$,
where $p_t$ is the transverse momentum in GeV/$c$.

Charged hadron identification is provided by $dE/dx$ measurements in the CDC,
an array of 1188 aerogel \v{C}erenkov counters (ACC), and a barrel-like array
of 128 time-of-flight scintillation counters (TOF); information from the three
subdetectors is combined to form a single likelihood ratio, which is then used
in kaon and pion selection. Electromagnetic showering particles are detected
in an array of 8736 CsI(Tl) crystals (ECL) that covers the same solid angle as
the charged particle tracking system. The energy resolution for electromagnetic
showers is $\sigma_E/E = (1.3 \oplus 0.07/E \oplus 0.8/E^{1/4})\%$, where $E$
is in GeV. Electron identification in Belle is based on a combination of
$dE/dx$ measurements in the CDC, the response of the ACC, and the position,
shape and total energy deposition (i.e., $E/p$) of the shower detected in the
ECL. The electron identification efficiency is greater than 92\% for tracks
with $p_{\rm lab}>1.0$~GeV/$c$ and the hadron misidentification probability is
below 0.3\%. The magnetic field is returned via an iron yoke that is
instrumented to detect muons and $K^0_L$ mesons. We use a GEANT-based Monte
Carlo (MC) simulation to model the response of the detector and determine its
acceptance~\cite{GEANT}.


\section{Event Reconstruction}

Charged tracks are selected with a set of track quality requirements based on
the number of CDC hits and on the distances of closest approach to the
interaction point. We also require that the track momenta transverse to the
beam be greater than 0.1~GeV/$c$ to reduce the low momentum combinatorial
background. Charged tracks that are positively identified as kaons, protons or
electrons are excluded. Since the muon identification efficiency and fake rate
vary significantly with the track momentum, we do not veto muons to avoid
additional systematic errors.

We identify $B$ candidates using two variables: the difference $\de$ between
the total reconstructed energy of a three-body combination and the nominal
beam energy in the c.m.\ frame and the beam constrained mass $\mb$. $\de$ is
calculated as $\de = E_B - E^*_{\rm beam} = 
\left (\sum_i\sqrt{c^2{\bf p}_i^2 + c^4m_i^2} \right ) - E^*_{\rm beam},$
where the summation is over all particles from a $B$ candidate; and ${\bf p}_i$
and $m_i$ are their c.m.\ three-momenta and masses, respectively. Since there
are no $\pi^0$'s or photons in the final state, the $\de$  width (with typical
value of 15~MeV) is governed by the track momentum resolution. The beam energy
spread is about 3~MeV and gives a negligible contribution to the total $\de$
width. The signal $\de$ shape is parametrized by a sum of two Gaussian
functions with a common mean. The $\de$ shape for the $\qqbar$ background is
parametrized by a linear function.
  The beam constrained mass variable $\mb$ is equivalent to the $B$ invariant
mass with the measured $B$ candidate energy $E_B$ replaced by the beam energy 
$E^*_{\rm beam}$:
$\mb = \frac{1}{c^2}\sqrt{E^{*2}_{\rm beam}-c^2{\bf P}_B^2} = 
       \frac{1}{c^2}\sqrt{E^{*2}_{\rm beam}-c^2(\sum_i {\bf p}_i)^2},$
where ${\bf P}_B$ is the $B$ candidate momentum in the c.m.\ frame. The average
$B$ meson momentum in the c.m.\ frame is about 0.34~\mom~which is much smaller
than its total energy. Thus, the uncertainty in the measured ${\bf P}_B$ gives
a small contribution to the $\mb$ width, which is dominated by the beam energy
spread. The $\mb$ width is about 3~MeV/$c^2$ and well described by a single
Gaussian function. The $\mb$ width in general, does not depend on the final
state (unless photons are included in the reconstructed final state).


\section{Background Suppression}

   There are two sources of the background: the dominant one is due to
$e^+e^-\to~\qqbar$ ($q = u, d, s$ and $c$ quarks) continuum events that have
a cross-section about three times larger than that for the
$e^+e^-\to\UFS\to\bbbar$; the other one originates from other $B$ meson decays.
The background from continuum events is suppressed using variables that
characterize the event topology. Since the two $B$ mesons produced from an
$\UFS$ decay are nearly at rest in the c.m.\ frame, their decay products are
uncorrelated and the event tends to be spherical. In contrast, hadrons from
continuum $\qqbar$ events tend to exhibit a two-jet structure. We use
$\theta_{\rm thr}$, which is the angle between the thrust axis of the $B$
candidate and that of the rest of the event, to discriminate between the two
cases. The distribution of $|\cos\theta_{\rm thr}|$, is strongly peaked near
$|\cos\theta_{\rm thr}|=1.0$ for $\qqbar$ events and is nearly flat for
$\bbbar$ events. We require $|\cos\theta_{\rm thr}|<0.80$ eliminating about
83\% of the continuum background and retaining 79\% of the signal events. For
further suppression of the continuum background we use a Fisher discriminant
formed from 11 variables: nine variables that characterize the angular
distribution of the momentum flow in the event with respect to the $B$
candidate thrust axis, the angle of the $B$ candidate thrust axis
with respect to the beam axis, and the angle between the $B$ candidate momentum
and the beam axis. The discriminant, $\cal{F}$, is the linear combination of
the input variables that maximizes the separation between signal and
background. The coefficients are determined using off-resonance data and a
large set of signal MC events. Use of the Fisher discriminant rejects about
89\% of the remaining continuum background with 53\% efficiency for the
signal. A more detailed description of the background suppression technique
can be found in Ref.~\cite{belle-khh2} and references therein.

The understanding of the background that originates from other $B$ meson decays
is of great importance in the study of charmless $B$ decays. We study the
$\bbbar$ related background using a large sample of MC generated $\bbbar$
generic events. We find that the dominant $\bbbar$ related background is due to
$B^0\to D^-\pi^+$, $D^-\to \ks\pi^-$ and due to $B^0\to J/\psi(\psi(2S))\ks$,
$J/\psi(\psi(2S))\to \mu^+\mu^-$ decays. We veto $B^0\to D^-\pi^+$ events by
requiring $|M(\ks\pi)-M_D|>100$~MeV/$c^2$. Modes with $J/\psi(\psi(2S))$
contribute due to muon-pion misidentification; the contribution from the
$J/\psi(\psi(2S))\to e^+e^-$ submode is found to be negligible after the
electron veto requirement. We exclude $J/\psi(\psi(2S))$ background by
requiring $|M(\pi^+\pi^-)_{\mu^+\mu^-}-M_{J/\psi}|>70$~MeV/$c^2$ and
$|M(\pi^+\pi^-)_{\mu^+\mu^-}-M_{\psi(2S)}|>50$~MeV/$c^2$, with a muon mass
assignment used here for the pion candidates. To suppress the background due
to $K/\pi$ misidentification, we also exclude candidates if the invariant mass
of any pair of oppositely charged tracks from the $B$ candidate is consistent
with the $D^+\to \ks K^+$ hypothesis within 15~MeV/$c^2$ ($\sim 2.5\sigma$),
regardless of the particle identification information. The most significant
background from charmless $B$ decays is found to originate from the decay
$B^0\to\eta'\ks$ followed by $\eta'\to\pi^+\pi^-\gamma$. There is also a
contribution from the two-body charmless decay $B^\pm\to\ks\pi^\pm$. Although
this background is shifted by about 0.2~GeV from the $\de$ signal region, it
is important to take it into account for correct estimation of the background
from continuum events.


\section{Three-body Signal Yields}
\label{sec:khh}

\begin{figure}[!t]
 \includegraphics[width=0.32\textwidth,height=45mm]{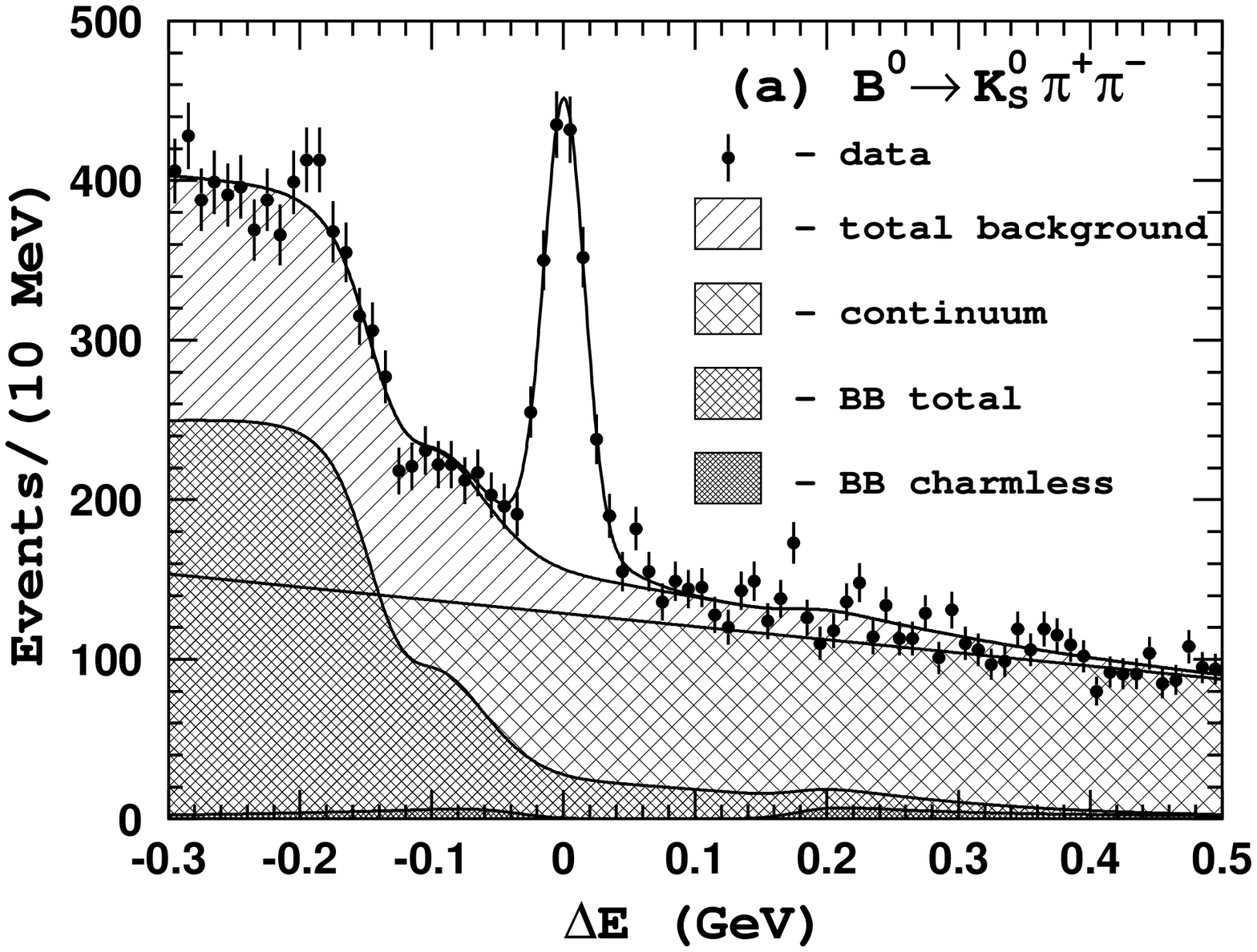} \hfill
 \includegraphics[width=0.32\textwidth,height=45mm]{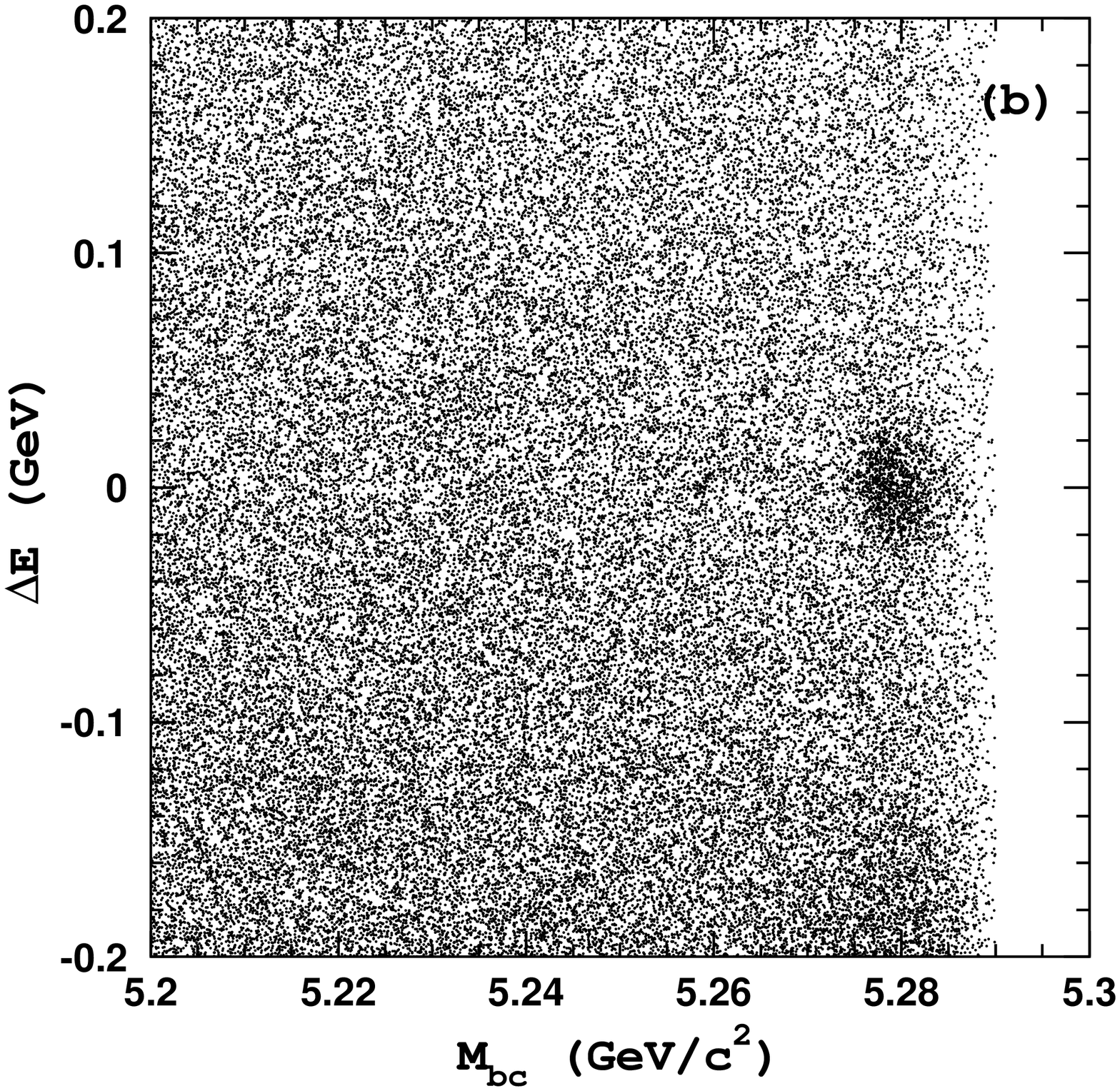} \hfill
 \includegraphics[width=0.32\textwidth,height=45mm]{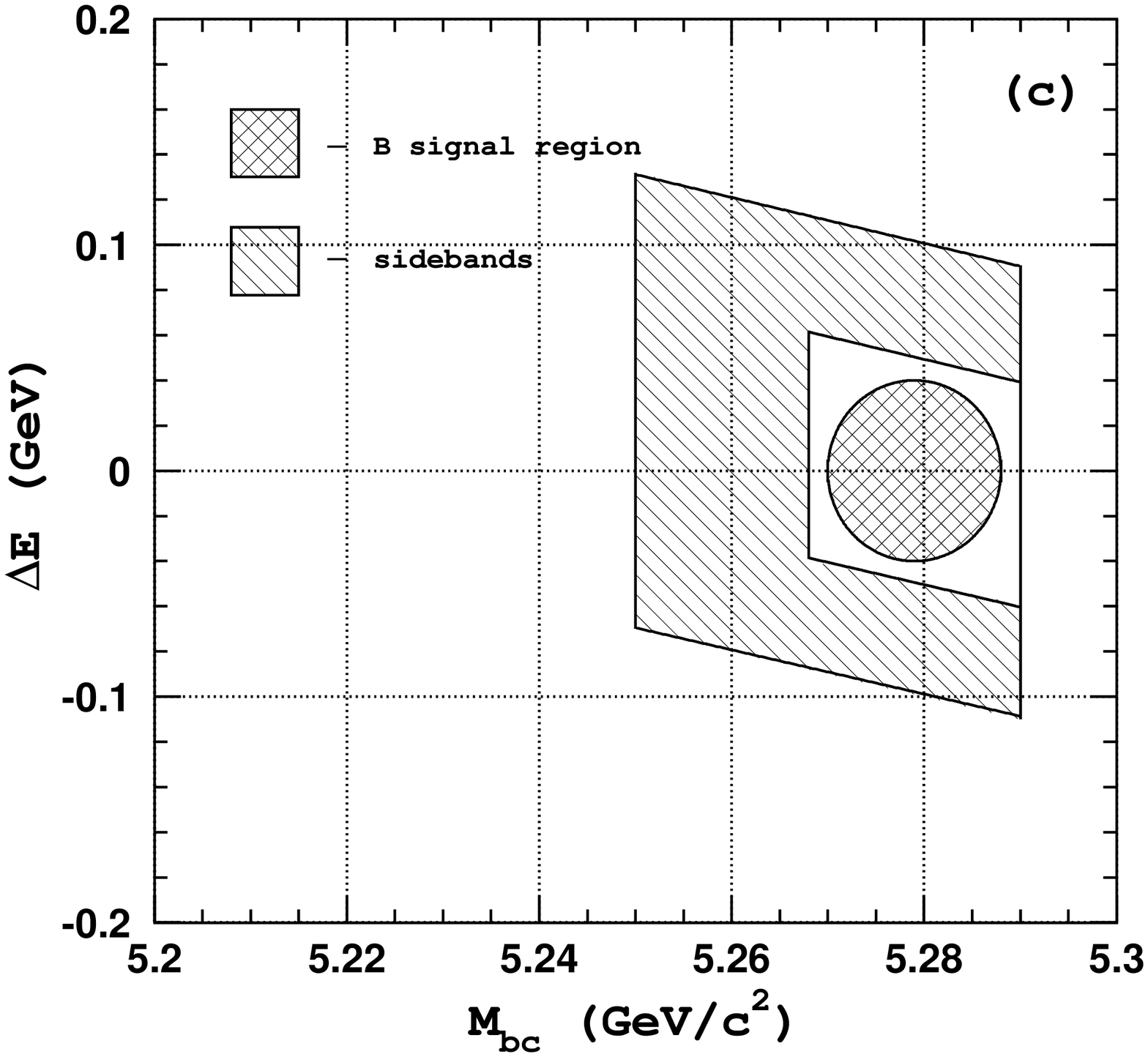}
 \caption{$\de$ distribution for the $\bnkpp$ candidate events with
          $|\mb-M_B|<7.5$~MeV/$c^2$. Points with error bars are data; the
          upper curve is the fit result; the hatched histogram is the
          background. \mbox{(b) Distribution} of $\de$ versus $\mb$ for
          the $\bnkpp$ candidates in data. \mbox{(c) Definition} of the
          $B$ signal and sideband regions in the $\mb-\de$ plane.}
 \label{fig:dE-Mbc}
\end{figure}

The $\de$ distributions for $\bnkpp$ candidates that pass all the selection
requirements are shown in Fig.~\ref{fig:dE-Mbc}(a), where clear peak in the
signal region is observed.  The two dimensional $\de$ versus $\mb$ plot is
shown in Fig.~\ref{fig:dE-Mbc}(b). In the fit to the $\de$ distribution we
fix the
shape and normalization of the charmless $\bbbar$ background components from
their measured branching fractions~\cite{PDG} and known number of produced
$\bbbar$ events. For the $\bbbar$ generic component we fix only the shape and
let the normalization float. The slope and normalization of the $\qqbar$
background component are free parameters. For signal we fix the width of the
second Gaussian function at 31.0~MeV and the fraction at 0.19 as determined
from MC simulation. The width of the main Gaussian is floating. The fit finds
$1229\pm62$ signal events. The sigma of the main Gaussian is
$15.3\pm0.9$~MeV. Results of the fit are shown in Fig.~\ref{fig:dE-Mbc}(a),
where different components of the background are shown separately for
comparison. There is a large increase in the level of the $\bbbar$ related
background in $\de<-0.15$~GeV region. This is mainly due to $B\to D\pi$,
$D\to K\pi\pi$ decay. This decay mode produces the same final state as the
studied process plus one extra pion that is not included in the energy
difference calculation. The semileptonic decays $B\to D^{(*)}\pi$,
$D\to K\mu\nu_\mu$ also contribute due to muon-pion misidentification. The
shape of the $\bbbar$ background is described well by MC simulation.

To examine possible quasi-two-body intermediate states in the observed $\bnkpp$
signal, we analyze the two-particle invariant mass spectra. To do so we define
the $B$ signal and sideband regions as illustrated in Fig.~\ref{fig:dE-Mbc}(c).
Defined in this way, the $\mb-\de$ sidebands are equivalent to the following
sidebands in terms of the three-particle invariant mass $M(K\pi\pi)$ and
three-particle momentum $P(K\pi\pi)$:
\[
0.05 {\rm ~GeV}/c^2 < |M(K\pi\pi)-M_B| < 0.10 {\rm ~GeV}/c^2;
~~~P(K\pi\pi) < 0.48 {\rm ~GeV}/c~~~~~~~  \nonumber
\]
and
\[
|M(K\pi\pi)-M_B| < 0.10 {\rm ~GeV}/c^2;
~~~0.48 {\rm ~GeV}/c < P(K\pi\pi) < 0.65 {\rm ~GeV}/c.~~~~~~~  \nonumber
\]
The $B$ signal region is defined as an ellipse around the $\mb$ and $\de$ mean
values:
\[
\frac{(\mb-M_B)^2}{(7.5~{\rm MeV}/c^2)^2} + \frac{\de^2}{(40~{\rm MeV})^2} < 1.
\]
The efficiency of the requirements that define the signal region is 0.923. The
total number of events in the signal region is 2207. The relative fraction of
signal events in the $B$ signal region is then determined to be
$0.521\pm0.025$.

\begin{figure}[!t]
  \centering
  \includegraphics[width=0.33\textwidth]{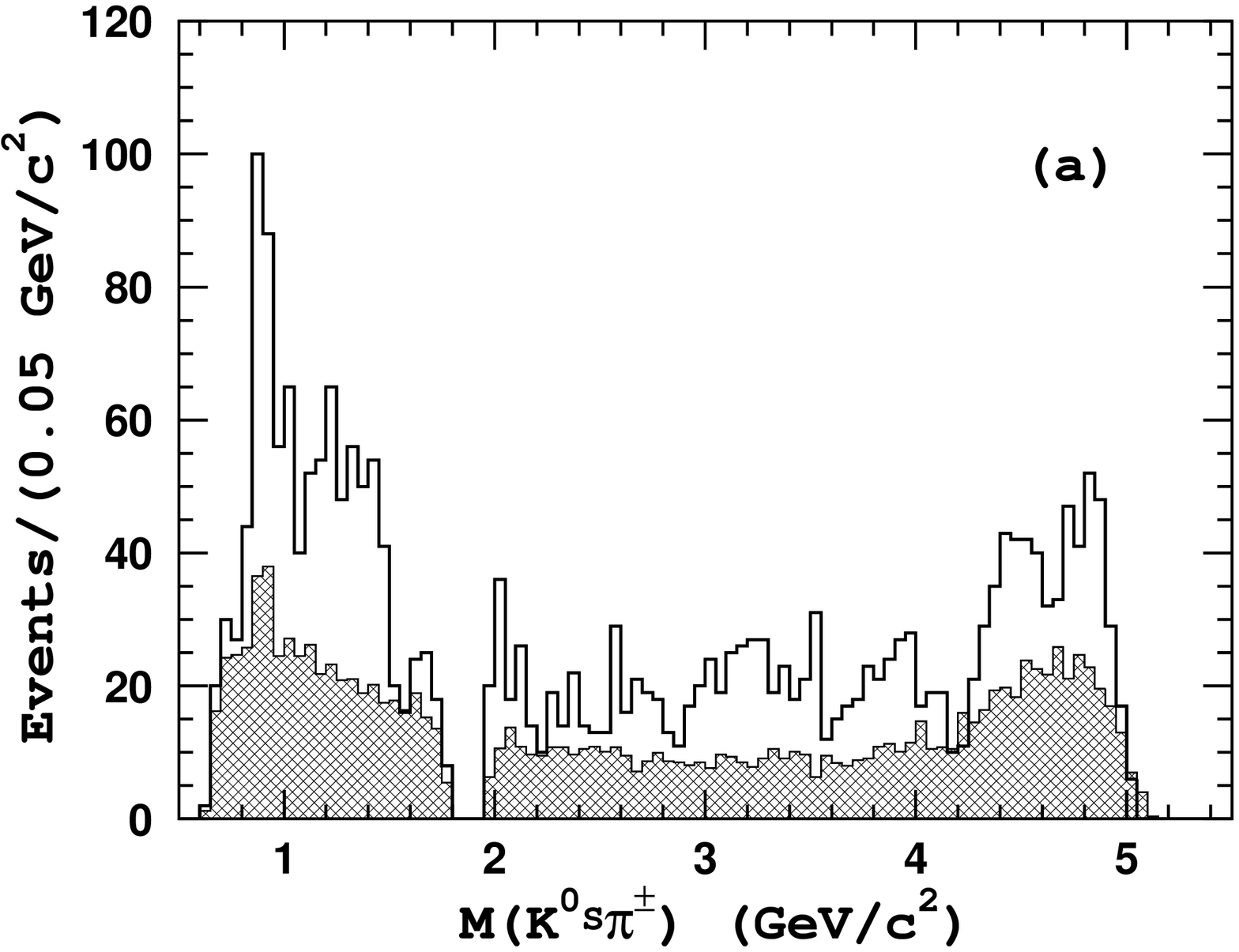}
\hspace*{-3mm}
  \includegraphics[width=0.33\textwidth]{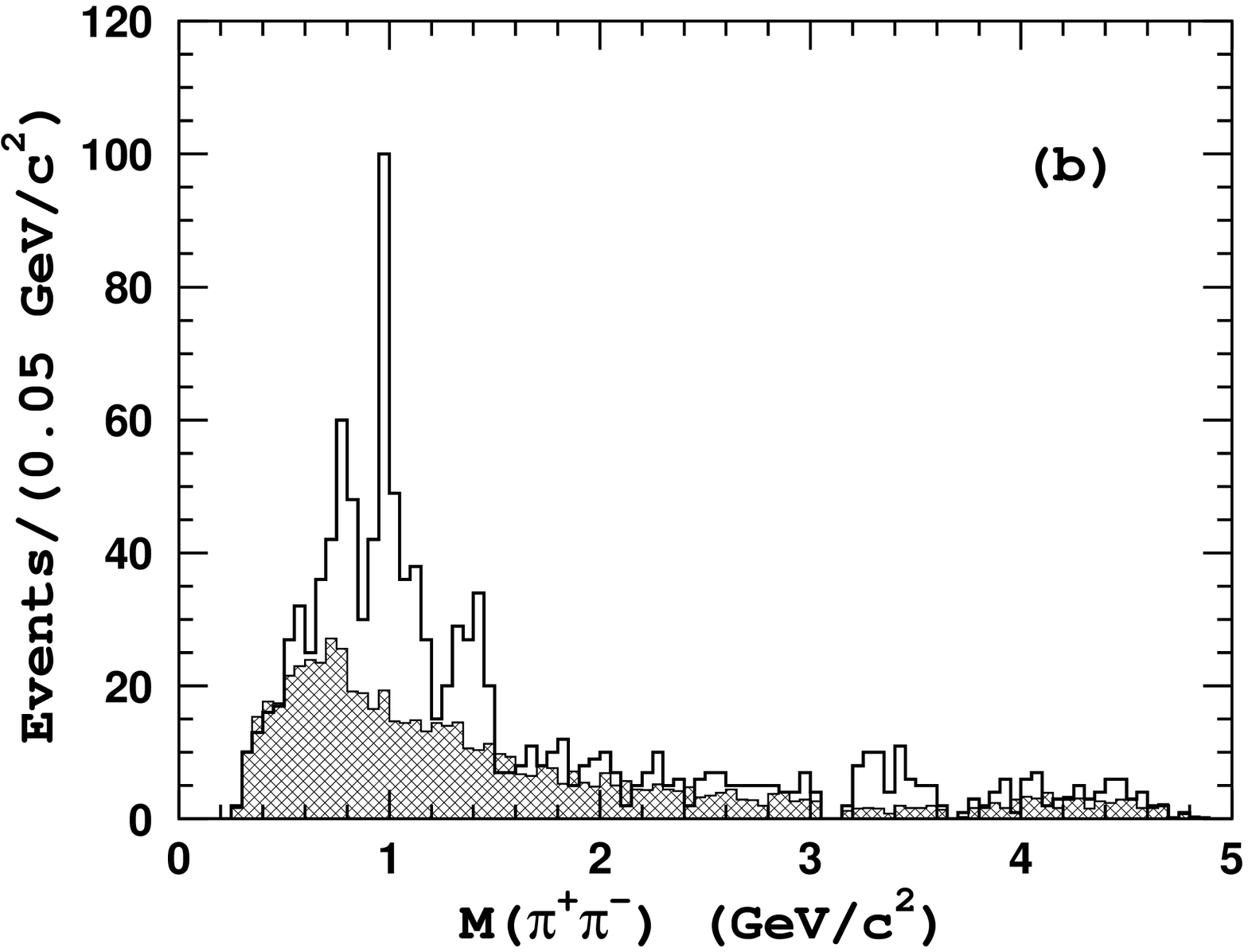}
\hspace*{-3mm}
  \includegraphics[width=0.33\textwidth]{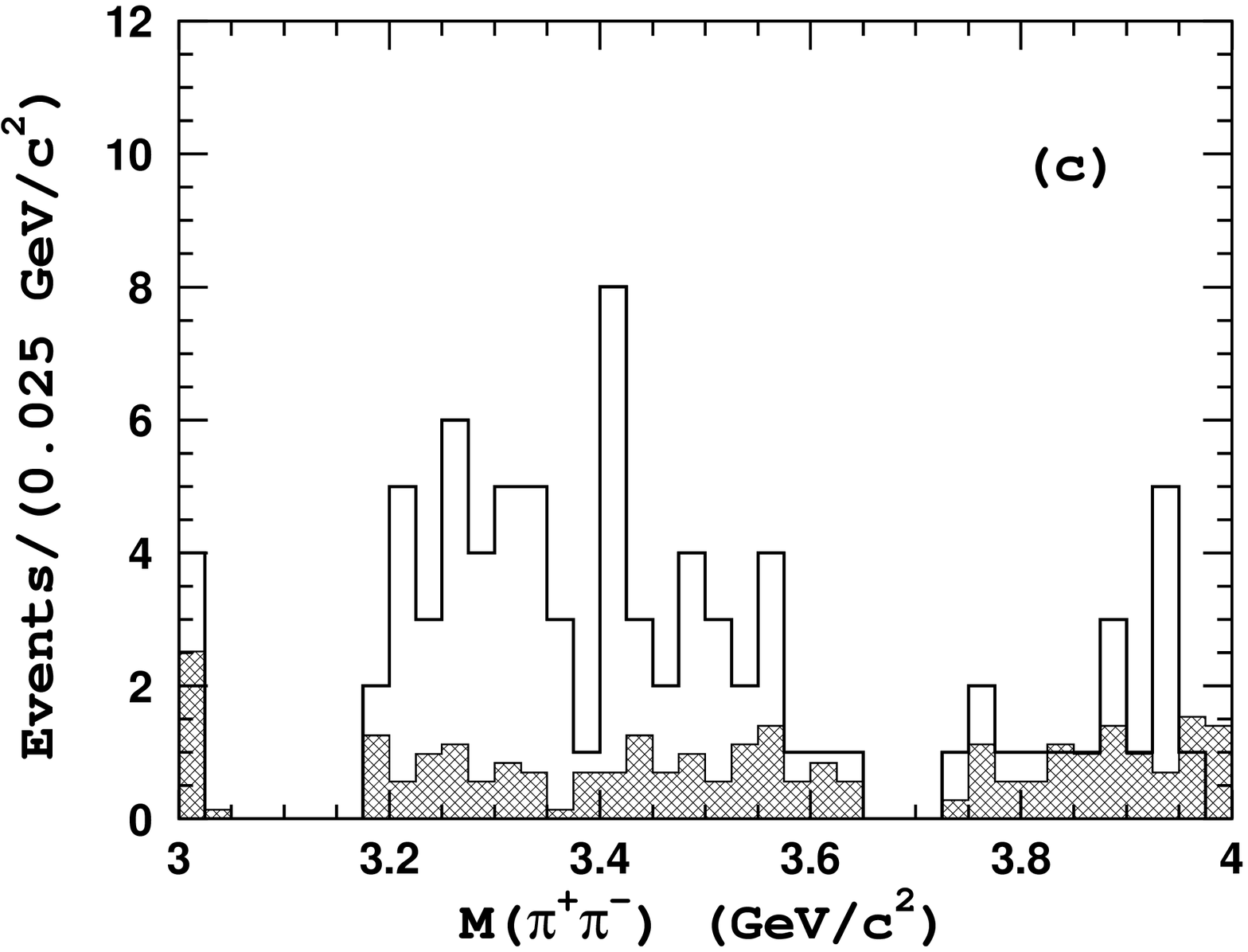}
  \caption{Two-particle invariant mass spectra for the $\bnkpp$ 
           candidates in the $B$ signal region (open histograms) and for
           background events in the $\de-\mb$ sidebands (hatched histograms). 
           (a) $M(\ks\pi^\pm)$ spectrum with $M(\pi^+\pi^-)>1.5$~\Mass
           (note that there are two entries per candidate in this plot);
           (b) $M(\pi^+\pi^-)$ with $M(\ks\pi^\pm)>1.5$~\Mass~ and
           (c) $M(\pi^+\pi^-)$ in the $\chic$ mass region with 
               $M(\ks\pi^\pm)>1.5$~\Mass.}
  \label{fig:kpp_hh}
\end{figure}

The $\kspi$ and $\pipi$ invariant mass spectra for $\bnkpp$ candidate events
in the $B$ signal region are shown as open histograms in
Fig.~\ref{fig:kpp_hh}. The hatched histograms show the corresponding spectra
for background events in the $\mb-\de$ sidebands, normalized to the estimated
number of background events. To suppress the feed-across between the $\pipi$
and $\kspi$ resonance states, we require the $\kspi$ ($\pipi$) invariant mass
to be larger than 1.5~\Mass ~when making the $\pipi$ ($\kspi$) projection.
The $\kspi$ invariant mass spectrum is characterized by a narrow peak around
0.9~GeV/$c^2$ which is identified as the $K^*(892)^\pm$ and a broad enhancement
around 1.4~\Mass. Possible candidates to assign to this enhancement are the
scalar $K^*_0(1430)^\pm$ and tensor $K^*_2(1430)^\pm$ resonances.
In the $\pipi$
invariant mass spectrum three distinct structures in the low mass region are
observed. The most prominent one is slightly below 1.0~\Mass~ and is consistent
with the $f_0(980)$. There is a clear indication for the $\rho(770)^0$ signal
to the left of the $f_0(980)$ peak.  Finally, there is a less prominent
structure between 1.2~\Mass ~and 1.5~\Mass. We cannot identify
unambiguously the resonance state that is responsible for such a structure;
possible candidates for a resonance state in this mass region might be
$f_0(1370)$, $f_2(1270)$ and perhaps $\rho(1450)$~\cite{PDG}. In what follows,
we refer to this structure as $f_X(1300)$. It is worth noting that both $K\pi$
and $\pi\pi$ two-body spectra in three-body $\bnkpp$ decays are similar to
those observed in charged $B$ meson decay to the three-body $\kpp$
final state~\cite{belle-khh3}. As we observe a clear $B^+\to\chic K^+$ signal
in the analysis of $\bckpp$ decay, we expect $B^0\to\chic K^0$ decay to occur
at a similar rate. However, the $\pipi$ mass spectrum for the $\chic$ region
shown in Fig.~\ref{fig:kpp_hh}(c) does not reveal the $\chic$ signal clearly.

From these qualitative considerations it is apparent that an amplitude analysis
is required for a more complete understanding of the individual quasi-two-body
channels that contribute to the observed three-body $\bnkpp$ signal.

\section{Amplitude Analysis}

In the preceding section we found that a significant fraction of the $\bnkpp$
signal can be assigned to quasi-two-body intermediate states. These resonances
will cause a non-uniform  distribution of events in phase space that can be
analyzed using the technique pioneered by Dalitz~\cite{dalitz}. Multiple
resonances that occur nearby in phase space will interfere providing an
opportunity to measure their amplitudes and relative phases. This in turn
allows us to deduce their relative fractions.

The amplitude analysis of $B$ meson three-body decays reported here is
performed by means of an unbinned maximum likelihood fit. As the unbinned
maximum likelihood fitting method does not provide a direct way to estimate
the quality of the fit, we need a measure to assess how well any given fit
represents the data. To do so the following procedure is applied. We first
subdivide the entire Dalitz plot into  1~\Masssq$\times$1~\Masssq~ bins. If
the number of events in the bin is smaller than $N_{\rm min}=16$ it is
combined with the adjacent bins until the number of events exceeds
$N_{\rm min}$. After completing this procedure, the entire Dalitz plot is
divided into a set of bins of varying size, and a $\chi^2$ variable for the
multinomial distribution can be calculated as
\begin{equation}
   \chi^2 = -2\sum^{N_{\rm bins}}_{i=1}n_i\ln\left(\frac{p_i}{n_i}\right),
\end{equation}
where $n_i$ is the number of events observed in the $i$-th bin, and $p_i$ is
the number of predicted events from the fit. For a large number of events
this formulation becomes equivalent to the usual one.
Since we are minimizing the unbinned likelihood function, our ``$\chi^2$''
variable does not asymptotically follow a $\chi^2$ distribution but it is
bounded by a $\chi^2$ variable with ($N_{\rm bins}-1$) degrees of freedom
and a $\chi^2$ variable with ($N_{\rm bins}-k-1$) degrees of
freedom~\cite{kendal}, where $k$ is the number of fit parameters. Because
it is bounded by two $\chi^2$ variables, it should be a useful statistic
for comparing the relative goodness of fits for different models. A more
detailed description of the technique is given in Ref.~\cite{belle-khh3}.


\begin{figure}[!t]
\includegraphics[width=0.49\textwidth]{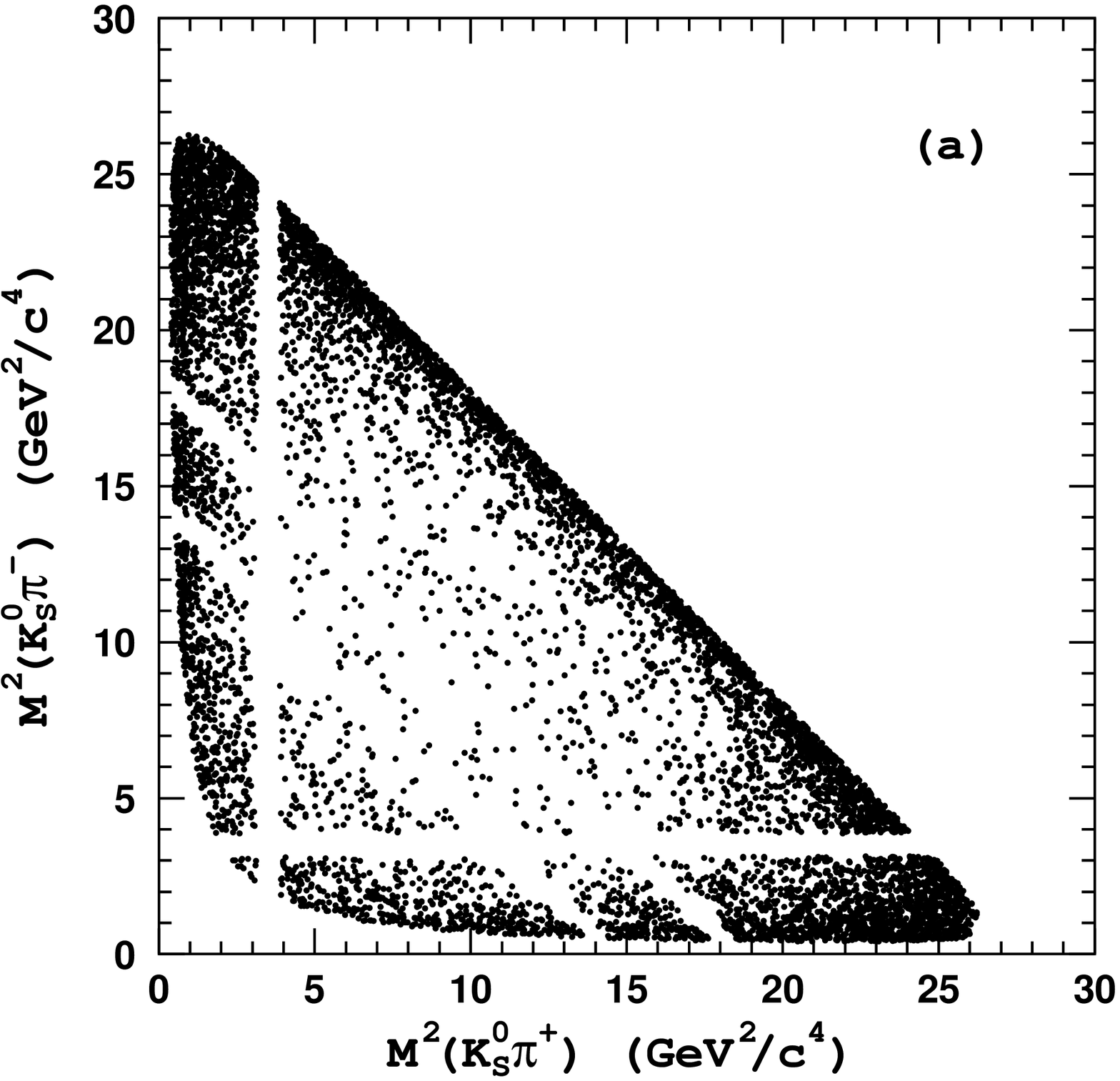} \hfill
\includegraphics[width=0.49\textwidth]{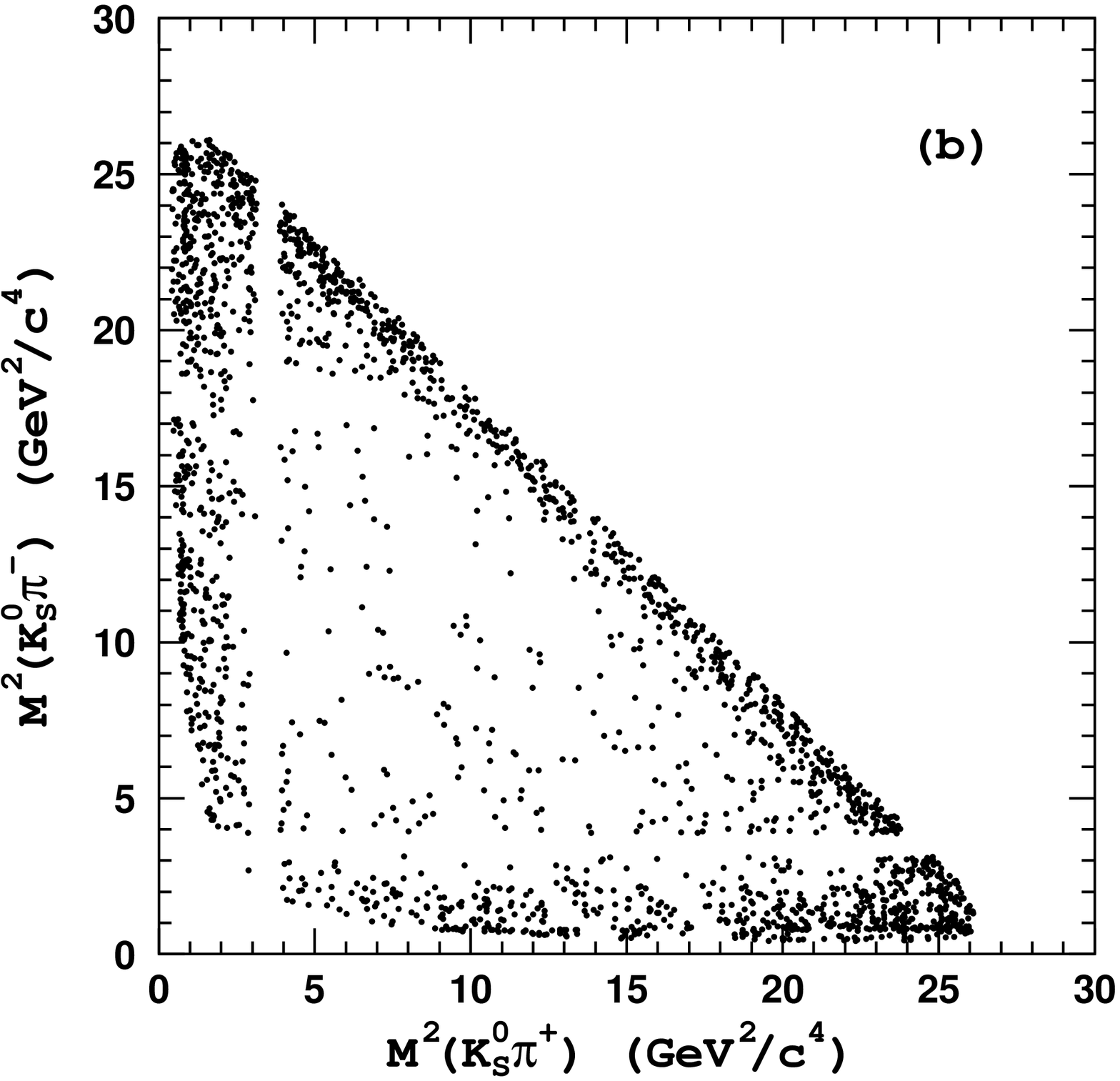}
  \caption{Dalitz plots for $\kspp$ candidates in the (a) $\de-\mb$ sidebands
           and (b) $B$ signal region.}
  \label{fig:khh-dp}
\end{figure}

\subsection{Fitting the Background Shape}
\label{sec:khh-bac}

Before fitting the Dalitz plot for events in the signal region, we need to
determine the distribution of background events. The background density
function is determined from an unbinned likelihood fit to the events in the
$\mb-\de$ sidebands defined in Fig.~\ref{fig:dE-Mbc}(c).
Figure~\ref{fig:khh-dp}(a) shows Dalitz distribution for 8159 sideband events.
This is about seven times the estimated number of background events in the
$B$ signal region.

\begin{figure}[!t]
  \centering
  \includegraphics[width=0.33\textwidth]{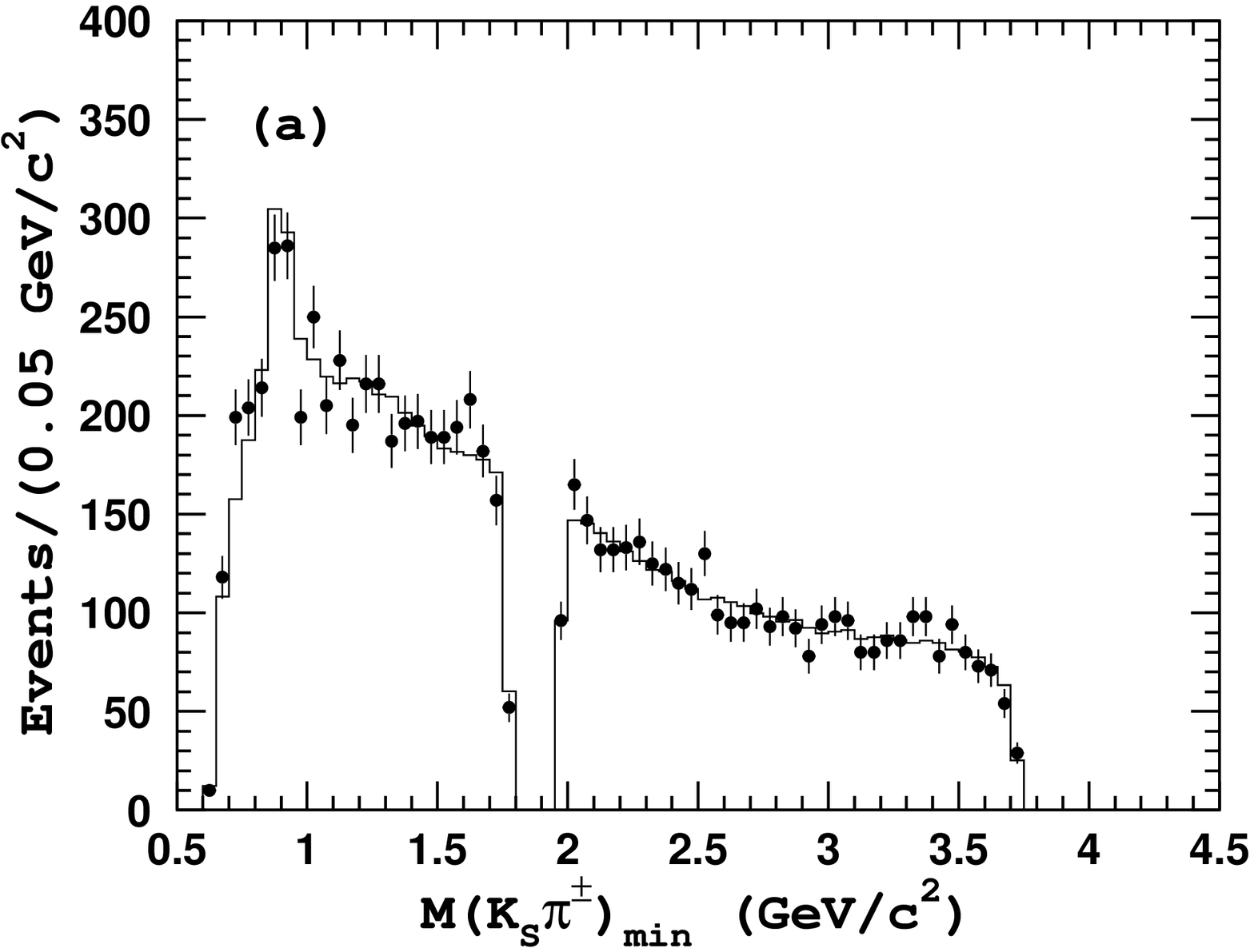} \hspace*{-3mm}
  \includegraphics[width=0.33\textwidth]{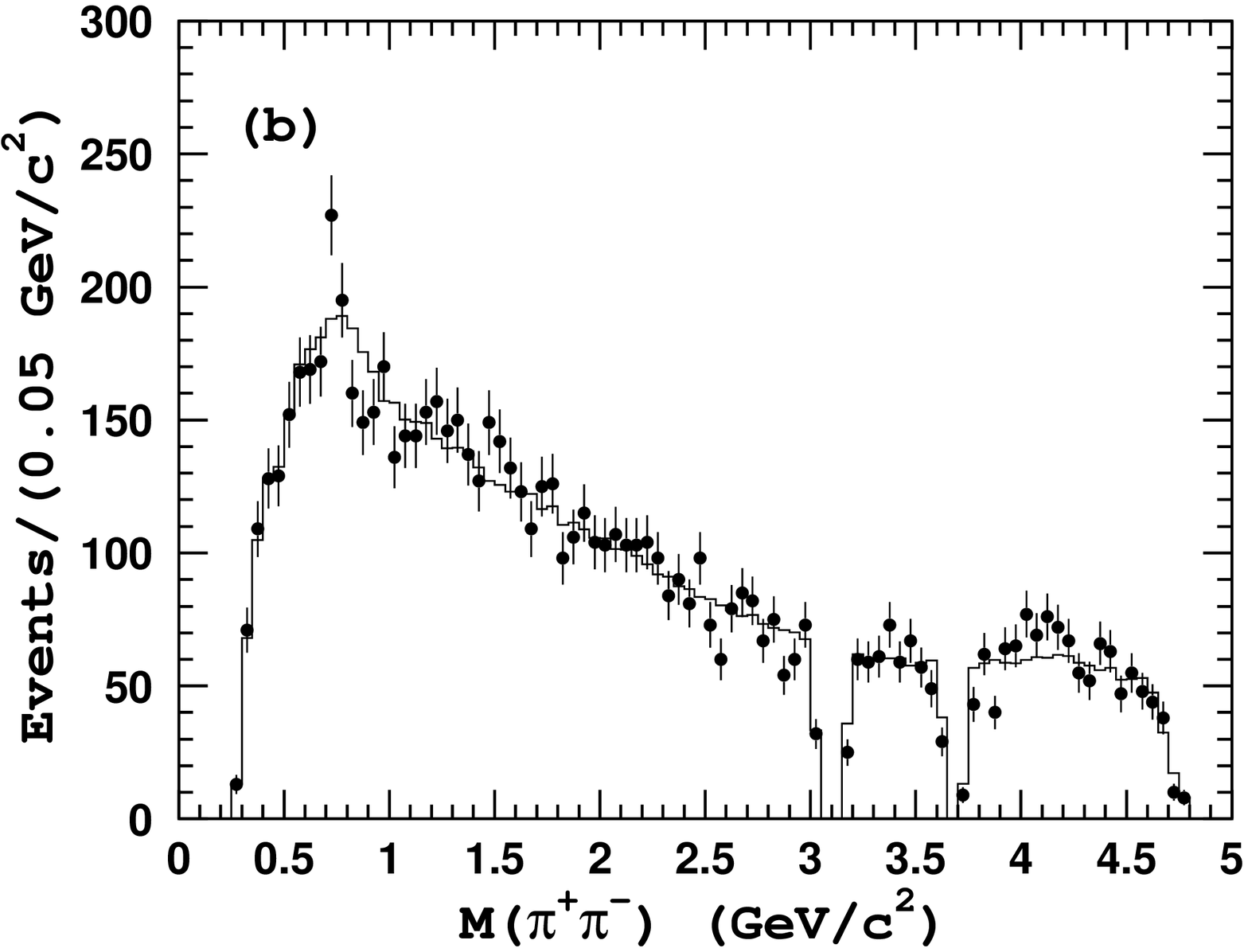} \hspace*{-3mm}
  \includegraphics[width=0.33\textwidth]{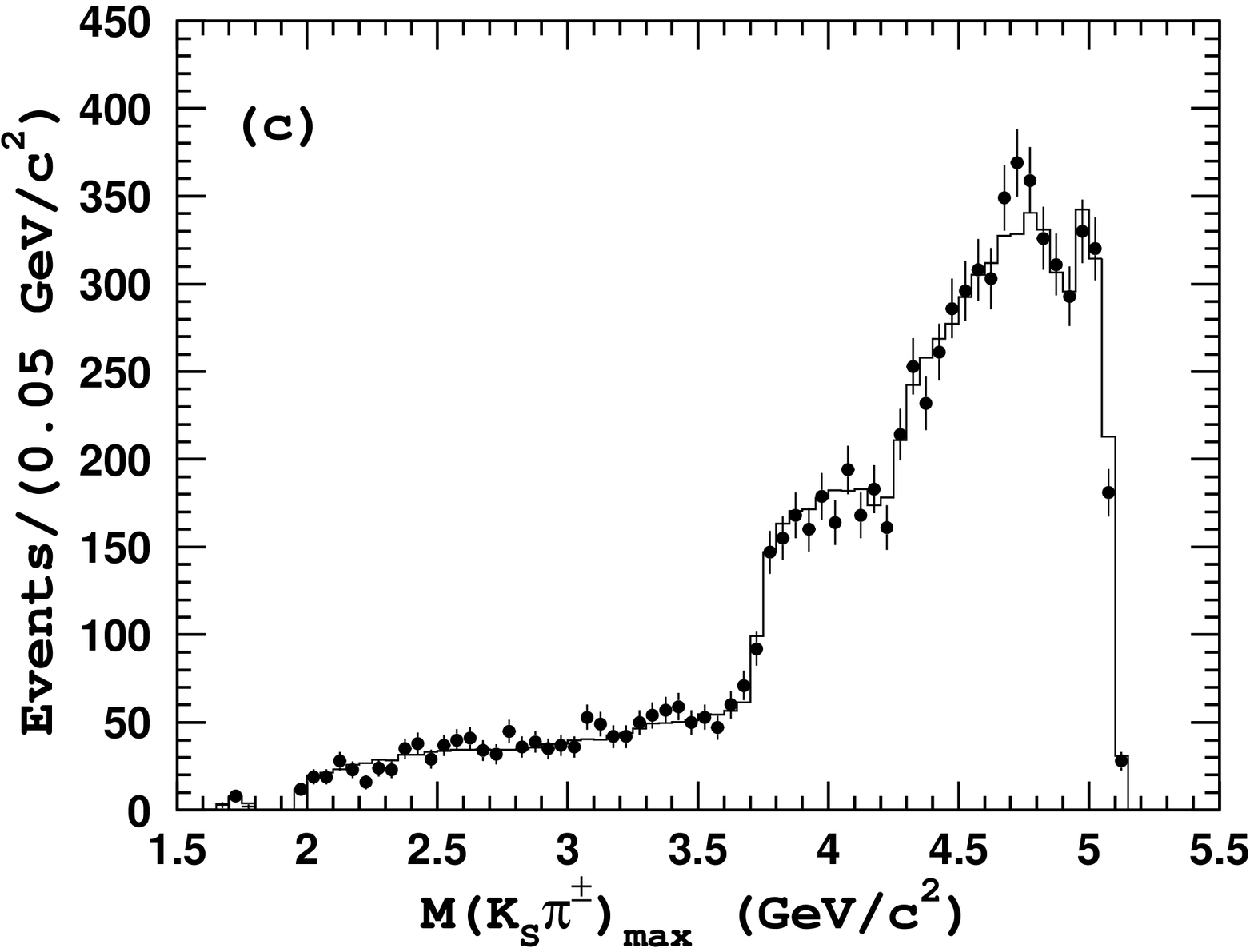}
  \caption{Results of the best fit to the $\kspp$ events in the $\de-\mb$ 
           sidebands shown as projections onto two-particle invariant mass
           variables. Points with error bars are data; histograms
           are fit results.}
\label{fig:khh_back}
\end{figure}

We use the following empirical parametrization to describe the distribution
of background events over the Dalitz plot in the $\kspp$ final state
\begin{eqnarray}
B(\sft,\sst) &=& \alpha_1(e^{-\beta_1s_{12}}+e^{-\beta_1s_{13}})
             ~+~ \alpha_2e^{-\beta_2s_{23}} \nonumber \\
             &+& \alpha_3(e^{-\beta_3s_{12}-\beta_4s_{23}}
             ~+~          e^{-\beta_3s_{13}-\beta_4s_{23}})
             ~+~ \alpha_4e^{-\beta_5(s_{12}+s_{13})} \nonumber \\
             &+& \gamma_1(|BW(K^*(892)^-)|^2+|BW(K^*(892)^+)|^2)
             ~+~ \gamma_2 |BW(\rho(770)^0)|^2,
\label{eq:kpp_back}
\end{eqnarray}
where $s_{12} \equiv M^2(\ks\pi^-)$,
$s_{13} \equiv M^2(\ks\pi^+)$, $s_{23}\equiv M^2(\pipi)$ and $\alpha_i$
($\alpha_1\equiv 1.0$), $\beta_i$ and $\gamma_i$ are fit parameters; $BW$
is a Breit-Wigner function.
The first three terms in Eq.~(\ref{eq:kpp_back}) are introduced to describe
the background enhancement in the two-particle low invariant mass regions.
This enhancement originates mainly from $e^+e^-\to\qqbar$ continuum
events. Due to the jet-like structure of this background, all three particles
in a three-body combination have almost collinear momenta. Hence, the
invariant mass of at least one pair of particles is in the low mass region.
In addition, it is often the case that two high momentum particles are
combined with a low momentum particle to form a $B$ candidate. In this
case there are two pairs with low invariant masses and one pair with high
invariant mass resulting in even  stronger enhancement of the background
in the corners of the Dalitz plot. This is taken into account by terms $4-6$
in Eq.~(\ref{eq:kpp_back}). To account for the contribution from real
$K^*(892)^\pm$ and $\rho(770)^0$ mesons, we introduce two more terms in
Eq.~(\ref{eq:kpp_back}), that are (non-interfering) squared Breit-Wigner
amplitudes, with masses and widths fixed at world average values~\cite{PDG}.
The projections of the data and fits for the background events are shown in
Figs.~\ref{fig:khh_back}. The $\chi^2/N_{\rm bins}$ value of the fit
is~$241.7/197$.

\subsection{Fitting the Signal}
\label{sec:kpp-sig}

\begin{figure}[!p]
  \centering
  \includegraphics[width=0.33\textwidth]{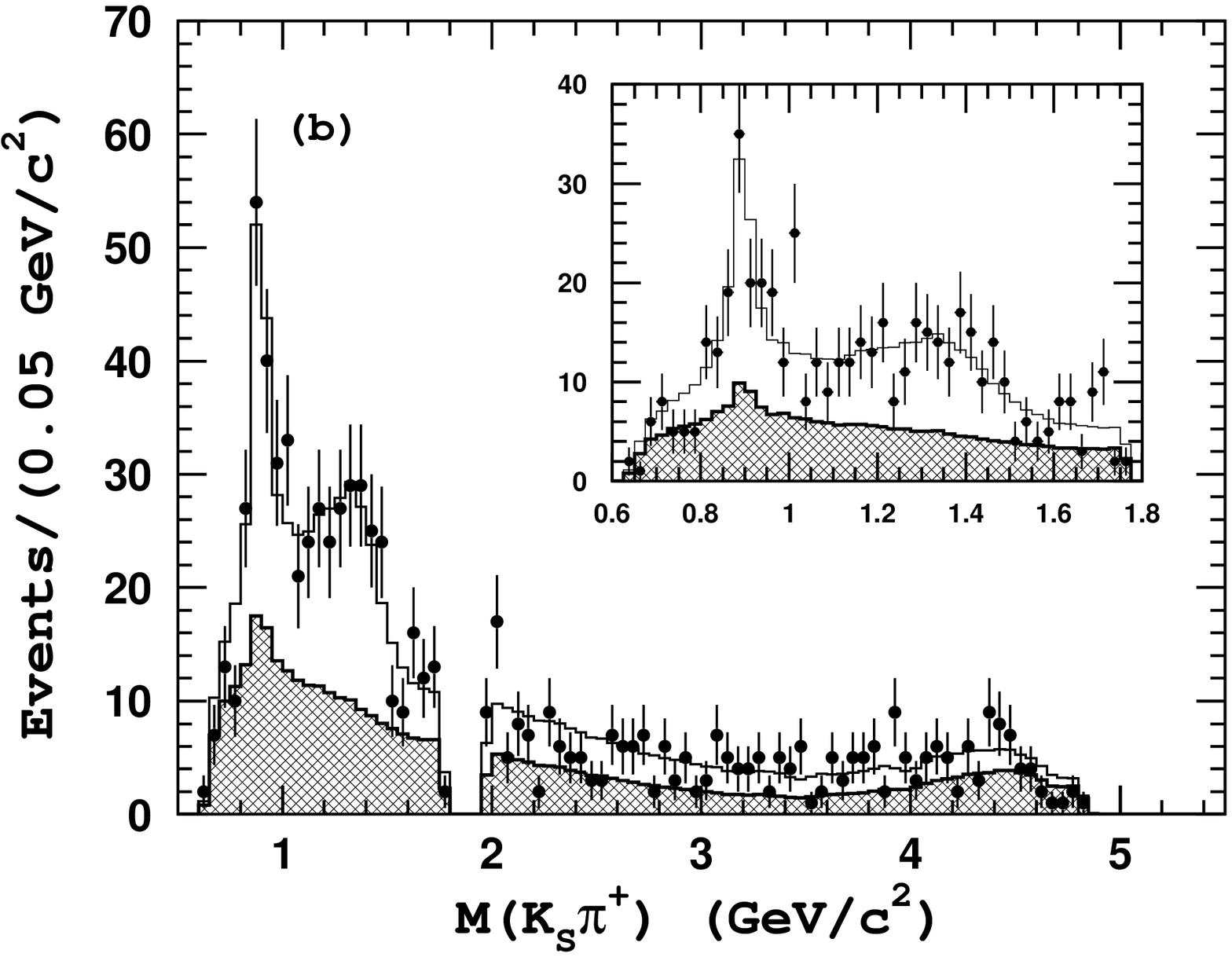} \hspace*{-2mm}
  \includegraphics[width=0.33\textwidth]{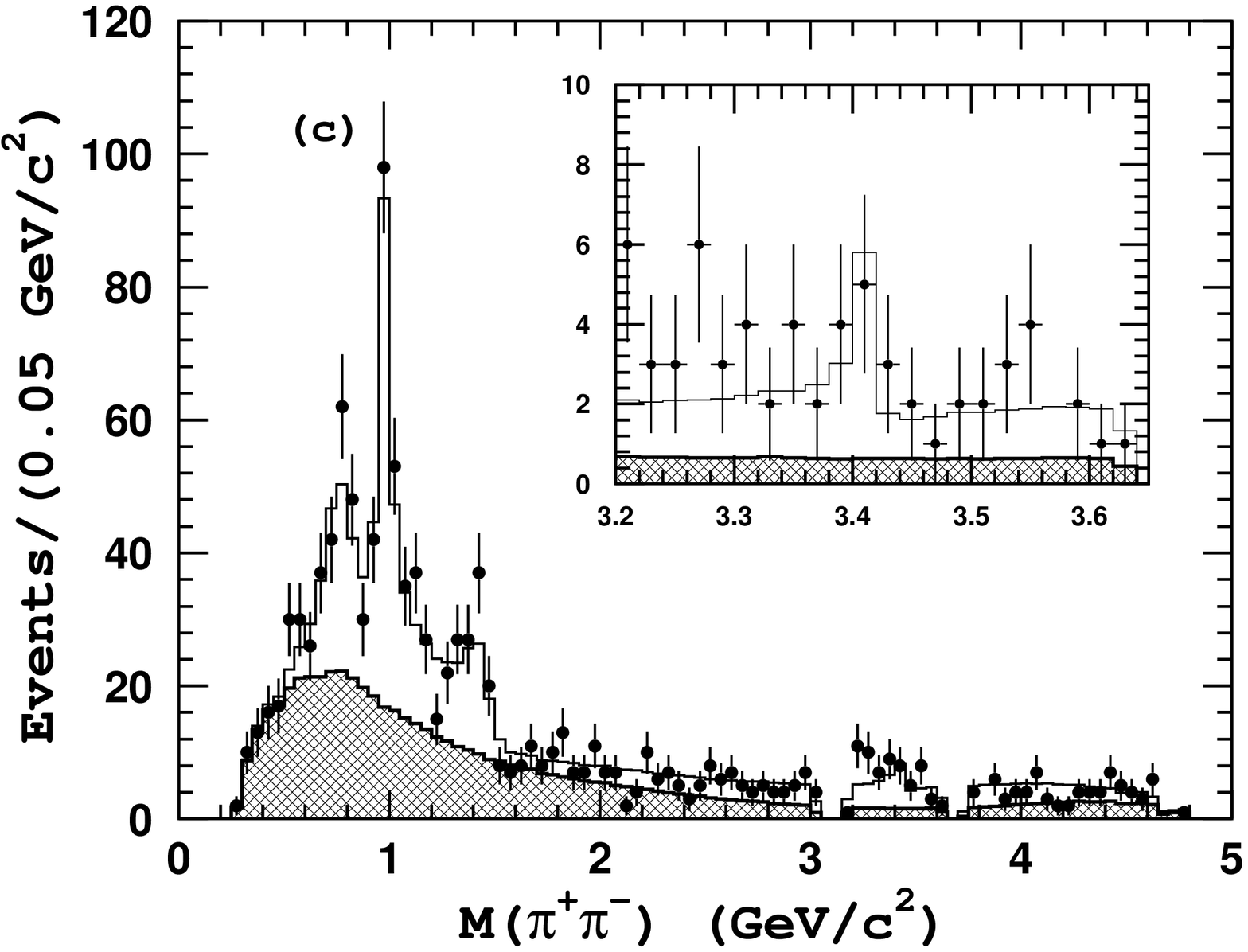} \hspace*{-2mm}
  \includegraphics[width=0.33\textwidth]{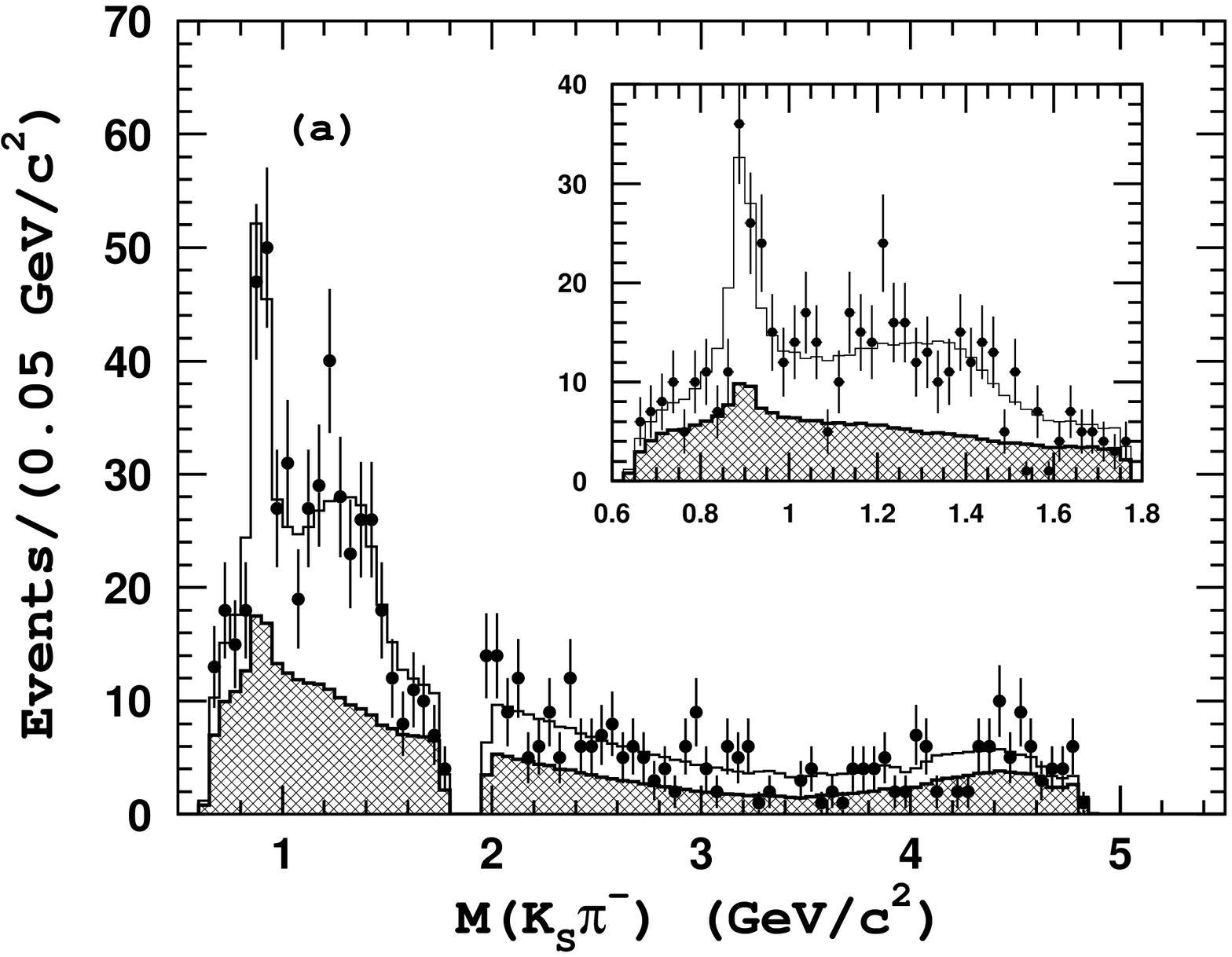}
  \caption{Results of the fit to $\kspp$ events in the signal region with
           the model~$\Kpp-$C$_0$. Points with error bars are data, the open
           histograms are the fit result and hatched histograms are the
           background components. Insets in (a) and (b) show the
           $K^*(892)-K_0^*(1430)$ mass region in 20~\mass~ bins; inset in (c)
           shows the $\chic$ mass region in 25~\mass~ bins.}
\label{fig:kpp-sig-fit}
\medskip
\medskip
  \centering
  \includegraphics[width=0.48\textwidth,height=55mm]{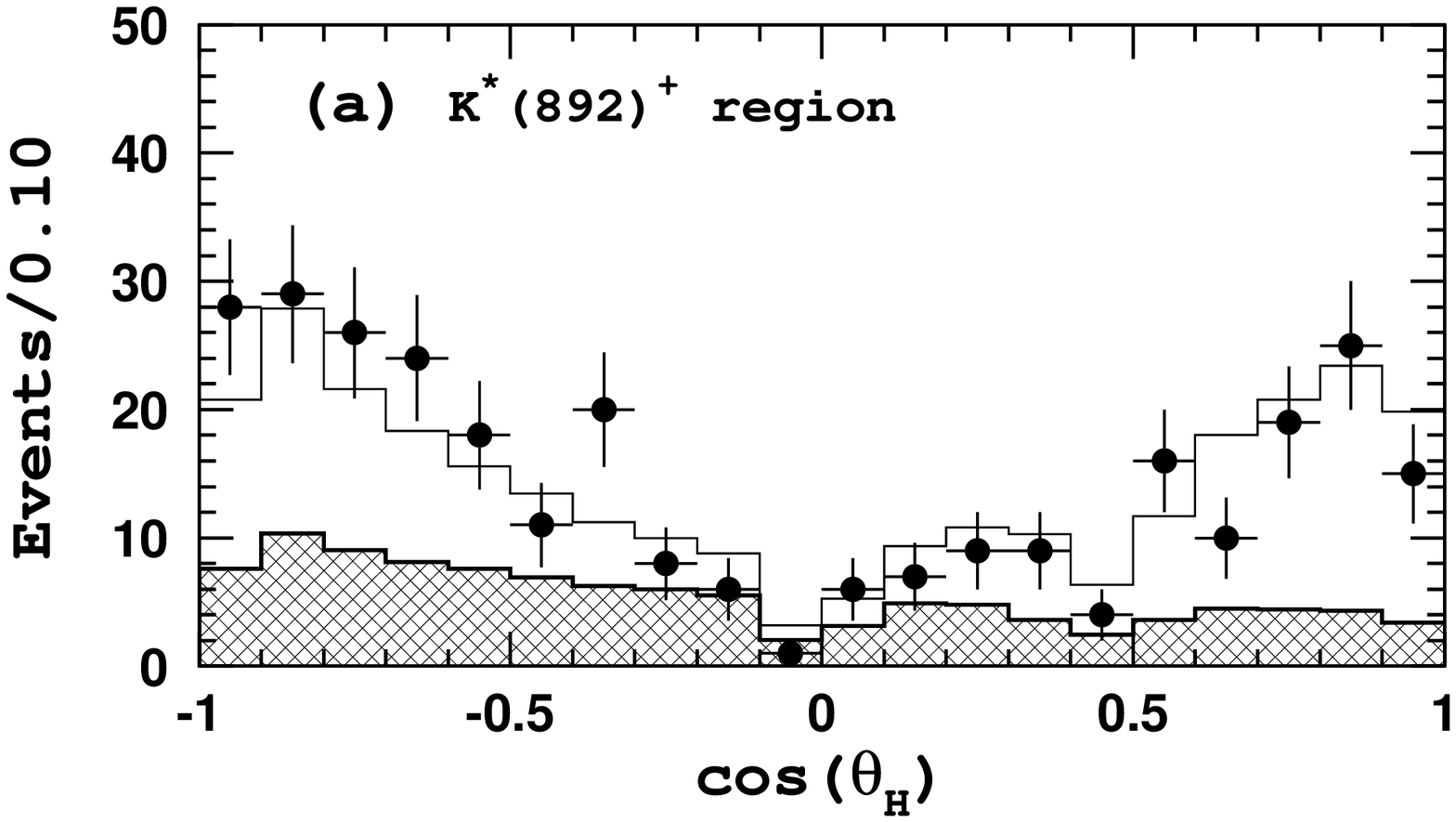} \hfill
  \includegraphics[width=0.48\textwidth,height=55mm]{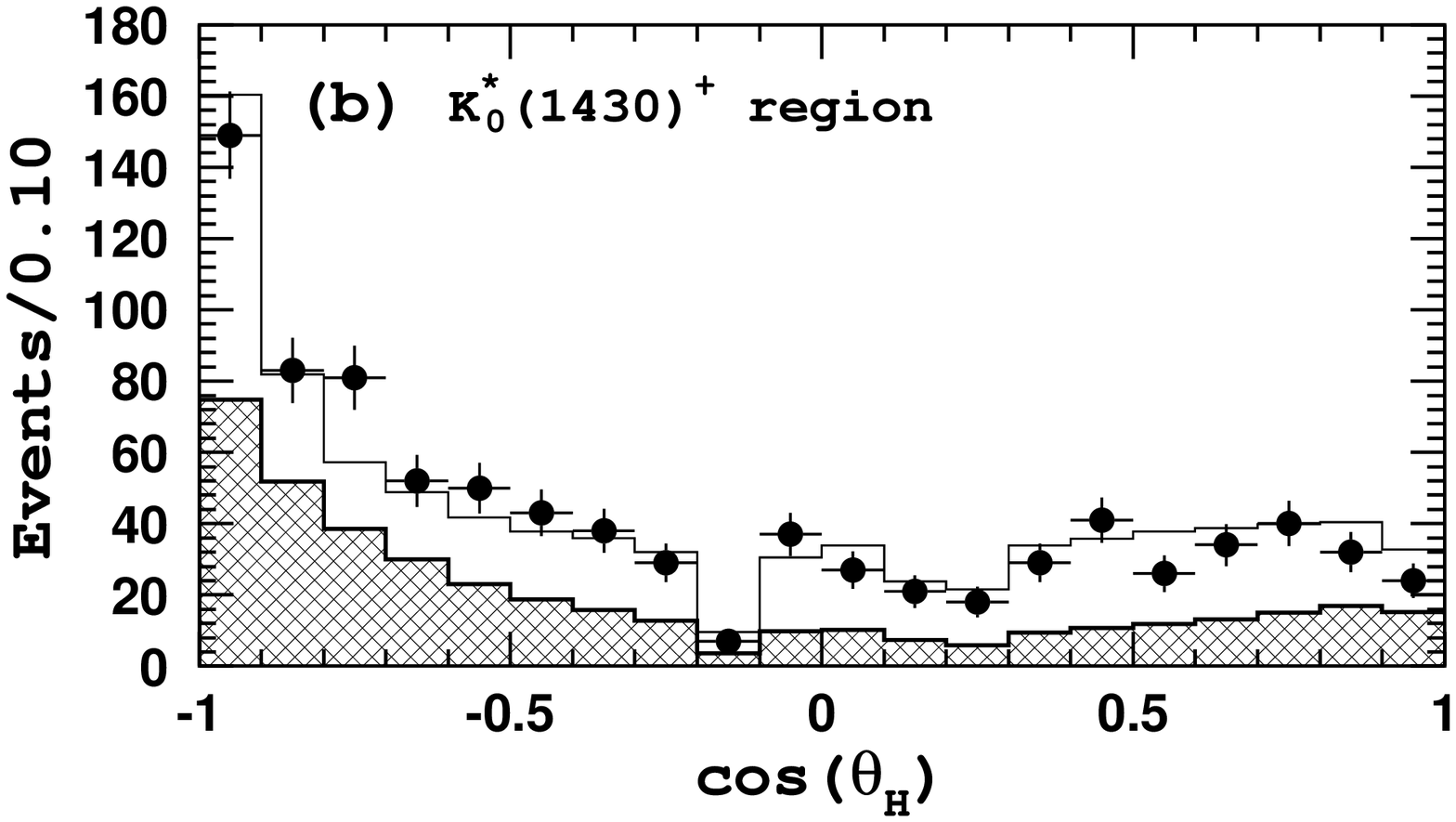} \\
  \includegraphics[width=0.48\textwidth,height=55mm]{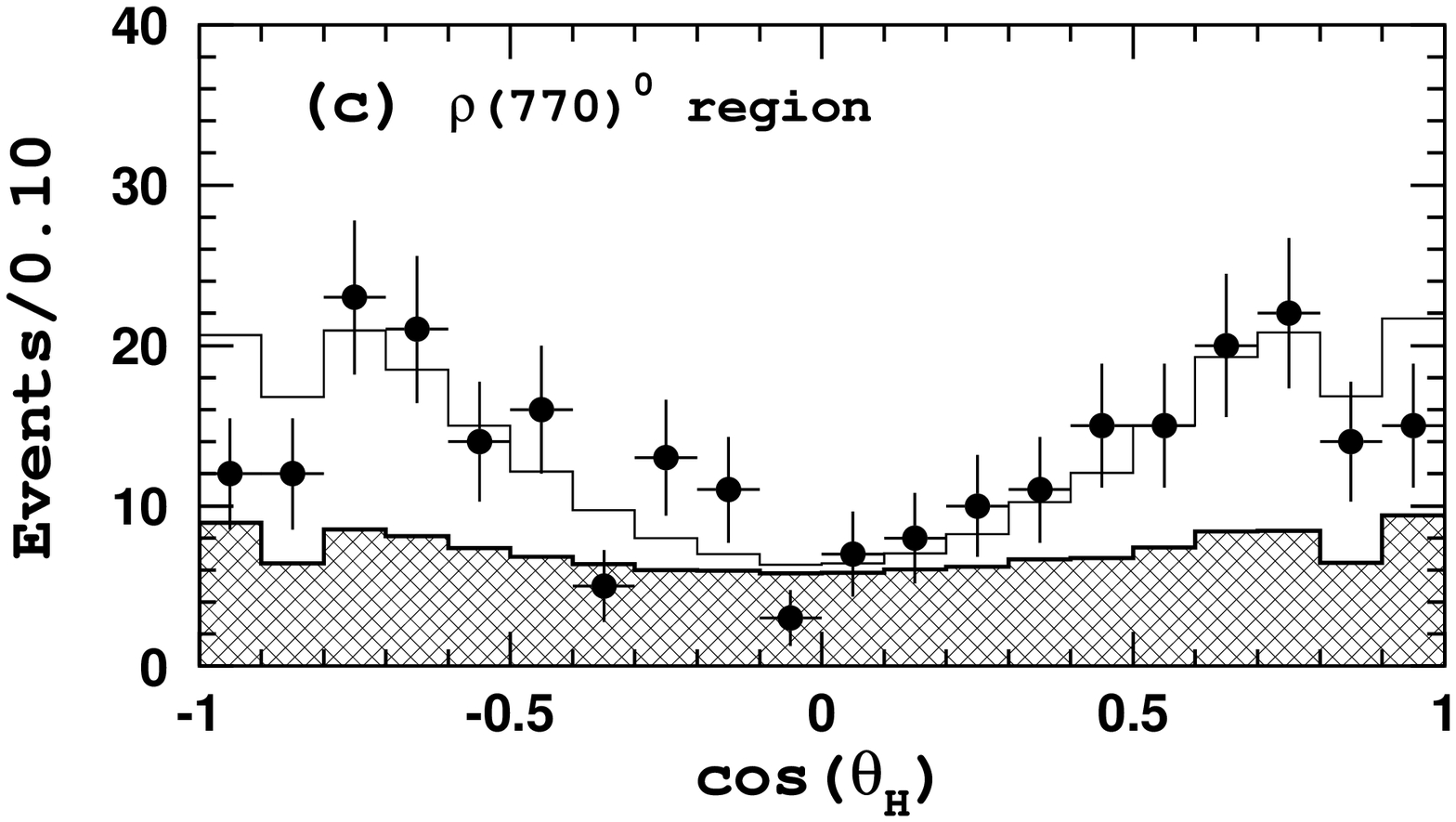} \hfill
  \includegraphics[width=0.48\textwidth,height=55mm]{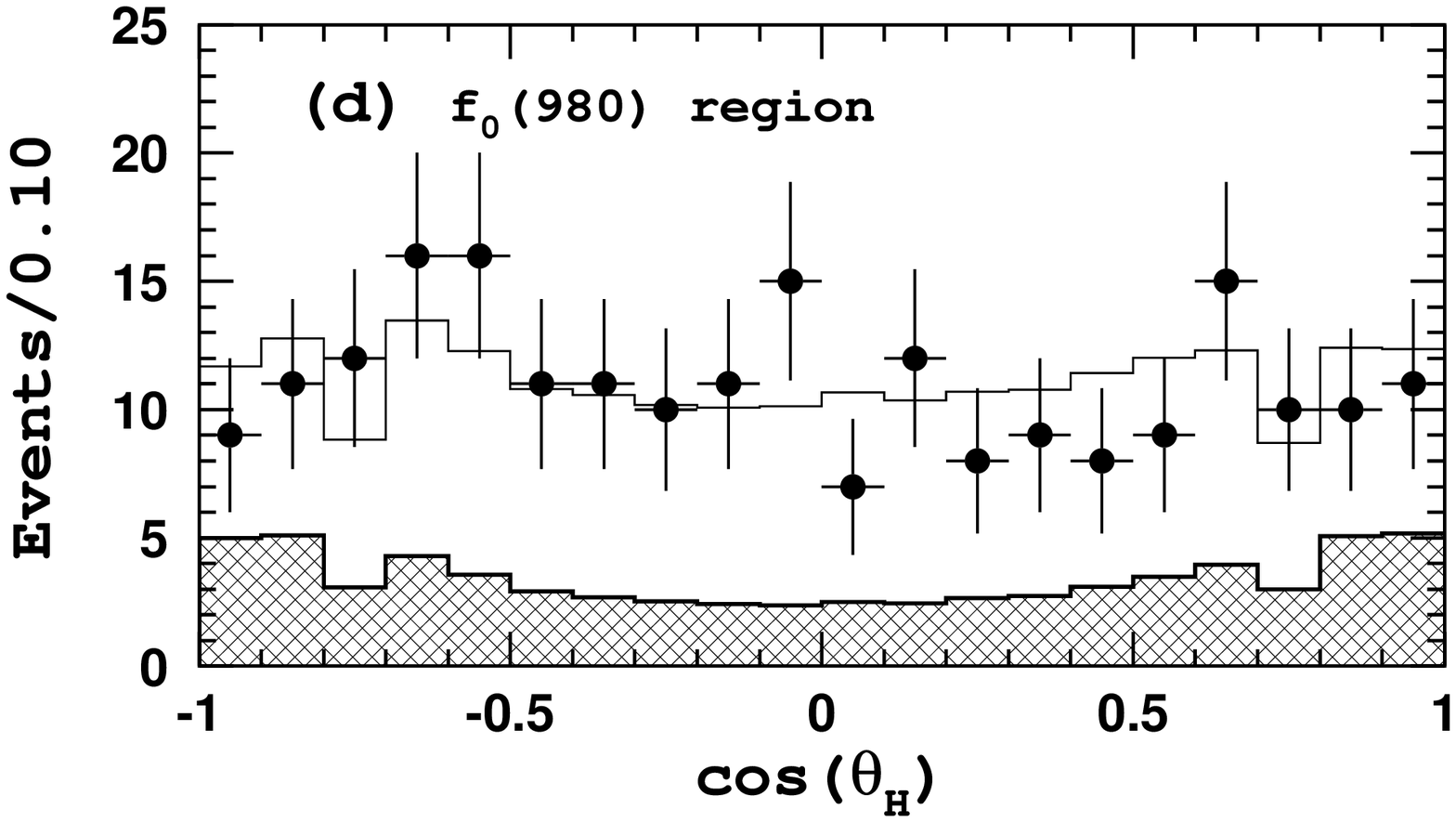}
  \caption{Helicity angle distributions for $\kspp$ events in different
           regions: \\
           (a)~$K^*(892)^+$ (0.82~\Mass~$<M(\ks\pi^\pm)<$~0.97~\Mass);\\
           (b)~$K^*_0(1430)^+$ (1.0~\Mass~$<M(\kcpi)<$~1.76~\Mass);\\
           (c)~$\rho(770)^0$ ($M(\pipi)<$~0.90~\Mass);\\
           (d)~$f_0(980)$ (0.90~\Mass~$<M(\pipi)<$~1.06~\Mass).
           Points with error bars are data, the open histogram is the fit
           result and the hatched histogram is the background component.
           Note that in plots (a) and (b) there are two entries per $B$
           candidate.}
\label{fig:kpp-heli}
\end{figure}

The Dalitz plot for $\kspp$ events in the signal region is shown in 
Fig.~\ref{fig:khh-dp}(b). There are 2207 events in the signal region that
satisfy all the selection requirements. In an attempt to describe all the
features of the $\kspi$ and $\pipi$ mass spectra mentioned above, we use
matrix element (refered to as model $\Kpp-$C$_J$) similar to those
constructed in the analysis of $\bckpp$ decay~\cite{belle-khh3}:
\begin{eqnarray}
{\cal{M}}_{C_J}(\kspp)
&=& a_{K^*}e^{i\delta_{K^*}}\Am_1(\pi^+\ks\pi^-|K^*(892)^+)
~+~ a_{K^*_0}e^{i\delta_{K^*_0}}\Am_0(\pi^+\ks\pi^-|K^*_0(1430)^+)\nonumber\\
&+& a_{\rho}e^{i\delta_{\rho}}\Am_1(\ks\pipi|\rho(770)^0)
~+~ a_{f_0}e^{i\delta_{f_0}}\Am_{\rm Flatte}(\ks\pipi|f_0(980)) \nonumber \\
&+& a_{f_X}e^{i\delta_{f_X}}\Am_J(\ks\pipi|f_X(1300))
~+~ a_{\chic}e^{i\delta_{\chic}}\Am_0(\ks\pipi|\chic) \nonumber \\
&+& \Am_{\rm nr}(\kspp),
\label{eq:kpp-modC}
\end{eqnarray}
where the subscript $J$ denotes the unknown spin of the $f_X(1300)$ resonance;
amplitudes $a_i$, relative phases $\delta_i$, are fit parameters. The masses
and widths of all resonances are fixed at either their world average
values~\cite{PDG} or at values determined from the analysis of $\bckpp$ decay
($f_0(980)$ and $f_X(1300)$). The $f_0(980)$ is parametrized with a coupled
channel Breit-Wigner function (Flatt\'e parametrization~\cite{Flatte}). For the
non-resonant amplitude~$\Am_{\rm nr}$ we use an empirical parametrization
\begin{equation}
{\cal A}_{\rm nr}(\kspp) =
      a^{\rm nr}_1e^{-\alpha{s_{13}}}e^{i\delta^{\rm nr}_1} 
    + a^{\rm nr}_2e^{-\alpha{s_{23}}}e^{i\delta^{\rm nr}_2}.
  \label{eq:kpp-non-res}
\end{equation}
Finally, as we currently do not perform the flavor analysis of the other $B$
meson, we cannot distinguish whether a $B$ or $\bar{B}$ meson decays to the
$\kspp$ final state. Thus the signal PDF is a non-coherent sum
\begin{equation}
   S_{C_J}(\kspp) = |{\cal{M}}_{C_J}(\ks\pi^+\pi^-)|^2 +
                    |{\cal{M}}_{C_J}(\ks\pi^-\pi^+)|^2.
\end{equation}

While fitting the data, we choose the $K^*(892)^+\pi^-$ signal as our reference
by fixing its amplitude and phase ($a_{K^*}\equiv 1$ and
$\delta_{K^*}\equiv 0$). The numerical values of the fit parameters are given
in Table~\ref{tab:kpp-fit-res}. Figure~\ref{fig:kpp-sig-fit} shows the fit
projections and the data. In addition, helicity angle distributions for several
regions are shown in Fig.~\ref{fig:kpp-heli}. The helicity angle is defined
as the angle between the direction of flight of the $\pi^+$ ($\pi^-$) in the
$\ks\pi^+$ ($\ks\pi^-$) rest frame and the direction of $\ks\pi$ system in the
$B$ rest frame. For the $\pi^+\pi^-$ system the helicity angle is defined with
respect to the positively charged pion. Gaps visible in
Figs.~\ref{fig:kpp-sig-fit} and~\ref{fig:kpp-heli} are due to vetoes applied
on invariant masses of two-particle combinations. All plots shown in
Figs.~\ref{fig:kpp-sig-fit} and~\ref{fig:kpp-heli} demonstrate
good agreement between data and the fit. We also try to fit the data assuming
$f_X(1300)$ is a vector (tensor) state. In this case we ascribe mass and width
of $\rho(1450)$ ($f_2(1270)$) from PDG~\cite{PDG} to it. If parametrized by a
single resonant state the best fit is achieved with a scalar assumption.

\begin{table}[!t]
\caption{Results of the fit to $\kspp$ events in the $B$ signal region
        with model $\Kpp-$C$_0$. The first quoted error is statistical
        and the second is the model dependent uncertainty.}
\medskip
\label{tab:kpp-fit-res}
\centering
  \begin{tabular}{@{\hspace{1mm}}l@{\hspace{2mm}}|@{\hspace{2mm}}c@{\hspace{2mm}}c@{\hspace{2mm}}c@{\hspace{2mm}}c@{\hspace{1mm}}} \hline \hline
     &                \multicolumn{4}{c}{Parameter}                    \\ 
\multicolumn{1}{@{\hspace{2mm}}c|@{\hspace{2mm}}}{Mode}  
     & Fraction, \% & Phase, $^\circ$ & Mass, GeV/$c^2$ & Width, GeV/$c^2$ \\
 \hline \hline  
 $K^*(892)^+\pi^-$    & $11.8\pm1.4\pm^{+0.9}_{-0.6}$  
                      & $0$ (fixed)    
                      & $0.89166$ (fixed)~\cite{PDG}
                      & $0.0508$  (fixed)~\cite{PDG} \\
 $K_0^*(1430)^+\pi^-$ & $64.8\pm3.9^{+1.6}_{-6.3}$
                      & $45\pm9^{+9}_{-13}$
                      & $1.412$ (fixed)~\cite{PDG}
                      & $0.294$ (fixed)~\cite{PDG}  \\
 $\rho(770)^0 K^0$    & $12.9\pm1.9^{+2.1}_{-2.2}$
                      & $-9\pm28^{+27}_{-13}$
                      & $0.7758$ (fixed)~\cite{PDG}
                      & $0.1503$ (fixed)~\cite{PDG}  \\
 $f_0(980)K^0$        & $16.0\pm3.4^{+1.0}_{-1.4}$
                      & $36\pm34^{+38}_{-21}$
                      & $0.950$ (fixed)~\cite{belle-khh4}
                      & $g_{\pi\pi}=0.23$ (fixed)~\cite{belle-khh4} \\
              &  & &  & $g_{KK}=0.73$ (fixed)~\cite{belle-khh4} \\
 $\chi_{c0}K^0$       & $0.43^{+0.42+0.02}_{-0.17-0.06}$
                      & $-99\pm37^{+8}_{-8}$
                      & $3.415$ (fixed)~\cite{PDG}
                      & $0.011$ (fixed)~\cite{PDG}  \\
 $f_X(1300)K^0$       & $3.68\pm2.16^{+0.53}_{-0.49}$
                      & $-135\pm25^{+24}_{-26}$
                      & $1.449$ (fixed)~\cite{belle-khh4}
                      & $0.126$ (fixed)~\cite{belle-khh4} \\
 Non-Resonant         & $41.9\pm5.1^{+1.4}_{-2.5}$
                      & $\delta^{\rm nr}_1=-22\pm8^{+6}_{-6}$
                      & $-$ & $-$ \\
                      &
                      & $\delta^{\rm nr}_2=175\pm30^{+54}_{-30}$ \\
 \hline  
 Charmless  Only      &  $99.3\pm0.4\pm0.1$ & $-$ & $-$ & $-$ \\
 \hline  \hline 
  \end{tabular}
\end{table}


\section{Model \& Systematic Uncertainties}

To estimate the model dependent uncertainty and test for the contribution of
other possible quasi-two-body intermediate states such as $K^*(1410)^+\pi^-$,
$K^*(1680)^+\pi^-$, $K^*_2(1430)^+\pi^-$ or $f_2(1270)K^0$, we include an
additional amplitude of either of these channels to model $\Kpp-$C$_0$ and
repeat the fit to data. For none of these channels is a statistically
significant signal found. We also use several alternative (though also
empirical) parametrizations of the non-resonant amplitude to estimate the
related uncertainty
\begin{itemize}
 \item{ $\Am_{\rm nr}(\kspp) =
          a^{\rm nr}_1e^{-\alpha \sft }e^{i\delta^{\rm nr}_1}$;}
 \item{ $\Am_{\rm nr}(\kspp) =
          a^{\rm nr}_1e^{-\alpha \sft }e^{i\delta^{\rm nr}_1}+
          a^{\rm nr}_2e^{-\alpha \sst}e^{i\delta^{\rm nr}_2}+
          a^{\rm nr}_3e^{-\alpha \sfs}e^{i\delta^{\rm nr}_3}$;}
 \item{ $\Am_{\rm nr}(\kspp)=
         \frac{a_1^{\rm nr}}{\sft^\alpha}e^{i\delta_1^{\rm nr}}+
         \frac{a_2^{\rm nr}}{\sst^\alpha}e^{i\delta_2^{\rm nr}}$.}
\end{itemize}

The dominant sources of systematic error are listed in Table~\ref{khh_syst}.
For the branching fraction of the three-body $\bnkpp$ decay, we estimate the
systematic uncertainty due to possible losses from the mass cuts used to
remove contributions from charmed particles by varying the relative phases
and amplitudes of the
quasi-two-body states within their errors. The systematic uncertainty due to
requirements on event shape variables is estimated from a comparison of the
$|\cos\theta_{\rm thr}|$ and ${\cal{F}}$ distributions for signal MC events
and $B^+\to\bar{D}^0\pi^+$ events in the data. The uncertainty from the
particle identification efficiency is estimated using pure samples of kaons and
pions from the $D^0\to K^-\pi^+$ decays, where the $D^0$ flavor is tagged using
$D^{*+}\to D^0\pi^+$ decays. We estimate the uncertainty due to the signal
$\de$ shape parameterization by varying the parameters of the fitting function
within their errors. The uncertainty in the background parameterization is
estimated by varying the relative fraction of the $\bbbar$ background component
and the slope of the $\qqbar$ background function within their errors.
The overall systematic uncertainty for the three-body branching fraction is
estimated to be $\pm7.7$\%.

\begin{table}[!t]
\centering
\caption{List of systematic errors (in percent) for the
         three-body $\bnkpp$ branching fraction.}
\medskip
\label{khh_syst}
  \begin{tabular}{lc}  \hline \hline
  Source~\hspace*{92mm} &  \multicolumn{1}{c}{Error}          \\
\hline 
 Charged track reconstruction &     $2.0$      \\
 PID                          &     $2.0$      \\
 $\ks$ reconstruction         &     $3.0$      \\
 Event Shape requirements     &     $2.5$      \\
 Signal yield  extraction     &     $5.4$      \\
 Model                        &     $2.2$      \\
 MC statistics                &     $1.0$      \\
 Luminosity measurement       &     $1.0$      \\
\hline
 Total                        &     $7.7$      \\
\hline \hline
  \end{tabular}
\end{table}


\section{Results}

In previous sections we determined the relative fractions of various
quasi-two-body intermediate states in the three-body $\bnkpp$ decay. To
translate these numbers into absolute branching fractions, we first need to
determine the branching fractions for the three body decay. To determine
the reconstruction efficiency for the $\bnkpp$ decay, we use MC simulation
where events are distributed over the phase space according to the matrix
elements of the model $\Kpp-$C$_0$. The corresponding reconstruction
efficiency is $6.71\pm0.03$\% (including $K^0\to\pi^+\pi^-$ fraction).

\begin{table}[t]
  \caption{Summary of branching fraction results. The first quoted error is
           statistical, the second is systematic and the third is the model
           error.}
  \medskip
  \label{tab:branch}
\centering
  \begin{tabular}{lcr} \hline \hline
\hspace*{20mm} Mode\hspace*{30mm} &
\hspace*{0mm} $\BF(B\to Rh)\times\BF(R\to hh)\times10^{6}$ &
\hspace*{10mm} $\BF(B\to Rh)\times10^{6}$  \\ \hline \hline
 $\kspp$  charmless total &
                        & $47.5\pm2.4\pm3.7$  \\
 $K^*(892)^+\pi^-$, $K^*(892)^+\to K^0\pi^+$
                        & $5.61\pm0.72\pm0.43^{+0.43}_{-0.29}$
                        & $8.42\pm1.08\pm0.65^{+0.64}_{-0.43}$     \\
 $K^*_0(1430)^+\pi^-$, $K^*_0(1430)^+\to K^0\pi^+$
                        & $30.8\pm2.4\pm2.4^{+0.8}_{-3.0}$
                        & $49.7\pm3.8\pm3.8^{+1.2}_{-4.8}$         \\
 $K^*(1410)^+\pi^-$, $K^*(1410)^+\to K^0\pi^+$
                        & $<3.8$ & $-$                             \\
 $K^*(1680)^+\pi^-$, $K^*(1680)^+\to K^0\pi^+$
                        & $<2.6$ & $-$                             \\
 $K^*_2(1430)^+\pi^-$, $K^*_2(1430)^+\to K^0\pi^+$
                        & $<2.1$ & $-$                             \\
 $\rho(770)^0K^0$, $\rho(770)^0\to\pi^+\pi^-$
                        & $6.13\pm0.95\pm0.47^{+1.00}_{-1.05}$
                        & $6.13\pm0.95\pm0.47^{+1.00}_{-1.05}$     \\
 $f_0(980)K^0$, $f_0(980)\to\pi^+\pi^-$
                        & $7.60\pm1.66\pm0.59^{+0.48}_{-0.67}$
                        & $-$                                      \\
 $f_2(1270)K^0$, $f_2(1270)\to\pi^+\pi^-$
                        & $<1.4$ & $-$                             \\
  Non-resonant
                        & 
                        & $19.9\pm2.5\pm1.5^{+0.7}_{-1.2}$         \\
\hline
 $\chic K^0$, $\chic\to\pi^+\pi^-$
                        & $<0.56$
                        & $<113$                                   \\
\hline \hline
  \end{tabular}
\end{table}

Results of the branching fraction calculations are summarized in
Table~\ref{tab:branch}. For final states where no statistically significant
signal is observed we calculate 90\% confidence level upper limits $f_{90}$
for their fractions. To determine the upper limit we use the following formula
\begin{equation}
  0.90 = \frac{\int_{0}^{f_{90}}G(a,s;x)dx}{\int_{0}^{\infty}G(a,s;x)dx},
\end{equation}
where $G(a,s;x)$ is a Gaussian function with mean $a$ and sigma $s$ which are
the measured mean value for the signal fraction and its statistical error.
To account for the systematic uncertainty we decrease the reconstruction
efficiency by one standard deviation.


\section{Discussion \& Conclusion}
\label{sec:discus}

With a $\lumi$ data sample collected with the Belle detector, an amplitude
(Dalitz) analysis of $B$ meson decays to three-body charmless $\kspp$ final
states is performed for the first time. Clear signals are observed in
the $B^0\to K^*(892)^+\pi^-$, $B^0\to K^*_0(1430)^+\pi^-$, 
$B^0\to\rho(770)^0K^0$ and $B^0\to f_0(980)K^0$
decay channels. The model uncertainty for these channels is small due to
their narrow width and, for vector-pseudoscalar decays, due to clear
signature because of the 100\% polarization of the vector meson. Among these
quasi-two-body channels the decay $B^0\to\rho(770)^0K^0$ is observed for the
first time.

The branching fraction measured for the decay $B^0\to K^*(892)^+\pi^-$ is in
agreement with results from the amplitude analysis of the three-body
$B^0\to K^+\pi^-\pi^0$ decay~\cite{babar-kpp,belle-kpp0}, where the
$K^*(892)^+$ is reconstructed in the $K^+\pi^0$ channel.

We also check possible contributions from $B^0\to K^*_2(1430)^+\pi^-$,
$B^0\to K^*(1410)^+\pi^-$, $B^0\to K^*(1680)^+\pi^-$ and $B^0\to f_2(1270)K^0$
decays. We find no statistically significant signal in any of these channels
and set 90\% confidence level upper limits for their branching fractions. In
the factorization approximation, charmless $B$ decays to final states with a
tensor state are expected to occur at the level of $\sim10^{-7}$~\cite{b2pt}.

We cannot identify unambiguously the broad structures observed in the
$M(\pipi)\simeq1.3$~\Mass~ mass region. If approximated by a single resonant
state, the best description is achieved with a scalar amplitude whose mass and
width are consistent with those for the $f_0(1370)$. Because of the large
uncertainty in $f_0(1370)$ parameters and its coupling to $\pipi$~\cite{PDG},
a more definite conclusion is not possible at present.


\section*{Acknowledgments}

   We thank the KEKB group for the excellent
   operation of the accelerator, the KEK Cryogenics
   group for the efficient operation of the solenoid,
   and the KEK computer group and the National Institute of Informatics
   for valuable computing and Super-SINET network support.
   We acknowledge support from the Ministry of Education,
   Culture, Sports, Science, and Technology of Japan
   and the Japan Society for the Promotion of Science;
   the Australian Research Council
   and the Australian Department of Education, Science and Training;
   the National Science Foundation of China under contract No.~10175071;
   the Department of Science and Technology of India;
   the BK21 program of the Ministry of Education of Korea
   and the CHEP SRC program of the Korea Science and Engineering Foundation;
   the Polish State Committee for Scientific Research under contract No.~2P03B 01324;
   the Ministry of Science and Technology of the Russian Federation;
   the Ministry of Education, Science and Sport of the Republic of Slovenia;
   the National Science Council and the Ministry of Education of Taiwan;
   and the U.S.\ Department of Energy.



\begin{thebibliography}{99} 

\bibitem{belle-khh3}
         A.~Garmash {\it et al.} (Belle Collaboration), \prd{71}{2005}{092003}.
\bibitem{belle-khh4}
         K.~Abe {\it et al.} (Belle Collaboration), hep-ex/0509001.
\bibitem{babar-ppp}
         B.~Aubert {\it et al.} (BaBar Collaboration), hep-ex/0408032.
\bibitem{babar-kpp}
         B.~Aubert {\it et al.} (BaBar Collaboration), hep-ex/0408073.
\bibitem{KEKB}S.~Kurokawa, \nima{499}{2003}{1}.
\bibitem{Belle}A.~Abashian {\it et al.}, \nima{479}{2002}{117}.
\bibitem{Ushiroda} 
 Y. Ushiroda (Belle SVD2 Group), \nima{511}{2003}{6}.
\bibitem{GEANT}
         R.Brun {\it et al.}, GEANT 3.21, CERN Report DD/EE/84-1, 1984.
\bibitem{belle-khh2}
         A.~Garmash {\it et al.} (Belle Collaboration), \prd{69}{2004}{012001}.
\bibitem{PDG} 
         S.~Eidelman {\it et al.} (Particle Data Group), \plb{592}{2004}{1}.
\bibitem{dalitz}
         R.H.~Dalitz,  Phil. Mag. {\bf 44}, 1068 (1953).
\bibitem{kendal}{M.G.~Kendall and A.~Stuart,
        {\it The Advanced Theory of Statistics}, 2nd ed.
        (Hafner Publishing, New York, 1968).}
\bibitem{Flatte}S.M.~Flatt\'e, \plb{63}{1976}{224}.
\bibitem{belle-kpp0}
         P.~Chang, {\it et al.} (Belle Collaboration), \plb{599}{2004}{148}.
\bibitem{b2pt}
C.S.~Kim, B.H.~Lim and S.~Oh, Eur. Phys. J. {\bf C22}, 683 (2002);\\
C.S.~Kim, J.P.~Lee and S.~Oh, \prd{67}{2003}{014002}.

\end{thebibliography}
\end{document}